\documentclass[twocolumn]{aastex61}
\usepackage{xspace}
\usepackage{latexsym}
\newcommand{\nraoblurb}{The National Radio Astronomy Observatory is
a facility of the National Science Foundation operated under cooperative
agreement by Associated Universities, Inc.}
\newcommand{\lsim}{\ensuremath{\,\lesssim\,}\xspace}
\newcommand{\hide}[1]{}
\newcommand{\gl}{\ensuremath{\ell}\xspace}
\newcommand{\gb}{\ensuremath{{\it b}}\xspace}

\newcommand{\kms}{\ensuremath{\,{\rm km\,s^{-1}}}\xspace}

\newcommand{\cm}{\ensuremath{\,{\rm cm}}\xspace}
\newcommand{\pc}{\ensuremath{\,{\rm pc}}\xspace}
\newcommand{\kpc}{\ensuremath{\,{\rm kpc}}\xspace}
\newcommand{\K}{\ensuremath{\,{\rm K}}\xspace}
\newcommand{\mK}{\ensuremath{\,{\rm mK}}\xspace}

\newcommand{\ghz}{\ensuremath{\,{\rm GHz}}\xspace}
\newcommand{\percc}{\ensuremath{\,{\rm cm^{-3}}}\xspace}

\newcommand{\degree}{\ensuremath{\,^\circ}\xspace}

\newcommand{\jy}{\ensuremath{\,{\rm Jy}}\xspace}

\newcommand{\mjy}{\ensuremath{\,{\rm mJy}}\xspace}

\newcommand{\rgal}{\ensuremath{\,R_{\rm Gal}}\xspace}   % Galactocentric radius
\newcommand{\hi}{{\rm H\,{\footnotesize I}}\xspace}
\newcommand{\hii}{{\rm H\,{\footnotesize II}}\xspace}

\newcommand{\hna}{\ensuremath{{\rm Hn}\alpha\xspace}}
\newcommand{\cor}{\ensuremath{^{\rm 13}{\rm CO}}\xspace}

\shorttitle{Diffuse Galactic HII Regions}
\shortauthors{Anderson et al.}

\begin{document}

\title{A Green Bank Telescope Survey of Large Galactic HII Regions}

\author{L.~D.~Anderson}
\affiliation{Department of Physics and Astronomy, West Virginia University, Morgantown WV 26506, USA}
\affiliation{Adjunct Astronomer at the Green Bank Observatory, P.O. Box 2, Green Bank WV 24944, USA}
\affiliation{Center for Gravitational Waves and Cosmology, West Virginia University, Chestnut Ridge Research Building, Morgantown, WV 26505, USA}
%\nocollaboration

\author{W.~P.~Armentrout}
\affiliation{Department of Physics and Astronomy, West Virginia University, Morgantown WV 26506, USA}
\affiliation{Center for Gravitational Waves and Cosmology, West Virginia University, Chestnut Ridge Research Building, Morgantown, WV 26505, USA}
%\nocollaboration

\author{Matteo~Luisi}
\affiliation{Department of Physics and Astronomy, West Virginia University, Morgantown WV 26506, USA}
\affiliation{Center for Gravitational Waves and Cosmology, West Virginia University, Chestnut Ridge Research Building, Morgantown, WV 26505, USA}
%\nocollaboration

\author{T.~M.~Bania}
\affiliation{Institute for Astrophysical Research, Department of Astronomy, Boston University, 725 Commonwealth Ave., Boston MA 02215, USA}
%\nocollaboration

\author{Dana~S.~Balser}
\affiliation{National Radio Astronomy Observatory, 520 Edgemont Road, Charlottesville VA, 22903-2475, USA}
%\nocollaboration

\author{Trey~V.~Wenger}
\affiliation{Astronomy Department, University of Virginia, P.O. Box 400325, Charlottesville, VA 22904-4325, USA}

%%%%%%%%%%%%%%%%%%%%%%%%%%%%%%%%%%%%%%%%%%%%%%%%%%
\begin{abstract}
  As part of our ongoing \hii\ Region Discovery Survey (HRDS), we
  report the Green Bank Telescope detection of 148 new angularly-large
  Galactic \hii\ regions in radio recombination line (RRL) emission.
  Our targets are located at a declination $\delta>-45\degree$, which
  corresponds to $266\degree > \ell > -20\degree$ at $b = 0\degree$.
  All sources were selected from the {\it WISE} Catalog of Galactic
  \hii\ Regions, and have infrared angular
  diameters $\ge 260\arcsec$.  The Galactic distribution of these
  ``large'' \hii\ regions is similar to that of the previously-known
  sample of Galactic \hii\ regions.
%  and we
%  propose that our sample is on average in a more evolved state
% compared with the general \hii\ region population.
  The large \hii\ region RRL line width and peak line intensity
  distributions are skewed toward lower values compared with that of
  previous HRDS surveys.
%  The low average peak line height reflects
%  the fact that these regions are low surface brightness.
  We discover 7 sources with extremely narrow RRLs $<10\,\kms$.  If
  half the line width is due to turbulence, these 7 sources have
  thermal plasma temperatures $<1100\,\K$.  These temperatures are
  lower than any measured for Galactic \hii\ regions, and the narrow
  line components may arise instead from partially ionized zones in
  the \hii\ region photo-dissociation regions.  We discover
  G039.515+00.511, one of the most luminous \hii\ regions in the
  Galaxy. We also detect the RRL emission from three \hii\ regions
  with diameters $>100\,\pc$, making them some of the physically
  largest known \hii\ regions in the Galaxy.
%  It is unclear how such low plasma temperatures can be
%  maintained within \hii\ regions without the collapse of the region
%  due to the ambient ISM pressure.
  This survey completes the HRDS \hii\ region census in the Northern
  sky, where we have discovered 887 \hii\ regions and more than
  doubled the previously-known census of Galactic \hii\ regions.
\end{abstract}


%%%%%%%%%%%%%%%%%%%%%%%%%%%%%%%%%%%%%%%%%%%%%%%%%%

\section{Introduction\label{sec:intro}}
\hii\ regions begin their lives as small zones of ionized gas
surrounding their central ionizing stars.  Because of the pressure
difference between the \hii\ region plasma and the neutral gas of the
ambient interstellar medium (ISM), they expand with time.  The total
mass of ionized gas in an \hii\ region is related to the stellar
output of ionizing photons and is therefore relatively constant
throughout the ionizing stars' main sequence lifetimes.  As they
evolve and increase in size, \hii\ region electron densities decrease
until they become large, diffuse nebulae \citep[e.g.,][]{dyson}.
%Despite their importance for understanding \hii\ region evolution, the
%\hii\ region luminosity function, and the origin of the warm ionized
%medium (WIM),

Their low infrared and radio surface brightness values make such
evolved nebulae difficult to observe.  Such large nebulae have been
detected in optical surveys \citep[e.g.][]{haffner03}, but there are
few radio studies, and even fewer studies at any other wavelength
dedicated to them.  Instead, much of the \hii\ region research over
the past $\sim$30~years has focused on young \hii\ regions, e.g.,
ultra-compact or hyper-compact nebulae.  Radio work on large
\hii\ regions as a class began and ended with \citet[][hereafter
  L96]{lockman96}, who observed 130 such nebulae in radio
recombination line (RRL) emission.
%(Some were previously known from optical studies, although
%Aside from this work, there are very few studies of large
%\hii\ regions \citep[although see][]{anderson09a, anderson09b}.
Our lack of knowledge about the large \hii\ region population means
that our sample of Galactic \hii\ regions remains incomplete.  Because
they have low surface brightnesses, such regions are also likely to be
missed in extragalactic studies.  This discrepancy in the \hii\ region
census may therefore impede our ability to accurately determine the
properties of Galactic star formation as traced by \hii\ regions.

Despite their low surface brightnesses, large \hii\ regions may be
among the most luminous in the Galaxy.  As part of the \hii\ Region
Discovery Survey \citep[HRDS;][]{bania10, anderson11}, we discovered
the ``G52L'' large \hii\ region, which had not yet been observed in
RRL emission \citep{bania12}.  We inferred the velocity of this nebula
through its association with compact \hii\ regions, together with \hi,
and \cor.
%The velocity of most large \hii\ region candidates cannot be so
%inferred because of the lack of associated compact \hii\ regions, \hi,
%and \co.
%G52L is illustrative, but it is an outlier.
This large region is 10\,kpc distant and has a 1.4\,GHz flux density
of $\sim11$\,Jy \citep{bania12}.  This
% actual numbers, excluding small regions are 10.9 Jy +/- 2.5
% =10^49.9 photons per second
implies an ionizing luminosity of $\sim10^{49.9}$\,s$^{-1}$,
equivalent to 2 main sequence O4 stars or $\sim10$ main sequence O7
stars \citep{sternberg03}.  Because it is relatively faint, G52L is
not in the \citet{murray10} catalog of large massive star forming
regions.  Its exclusion from their catalog hints that other, yet to be
discovered, luminous large \hii\ regions such as G52L may
make important contributions to the total ionization of the Galaxy.

Large \hii\ regions may also play a key role in maintaining the warm
ionized medium (WIM), the origins of which are still debated.  In
  many cases, we observe that the photo-dissociation regions (PDRs) of
  large \hii\ regions are fragmentary, which indicates that they may be
  leaking a significant number of ionizing photons into the ISM.
Leaking photons from the OB stars powering \hii\ regions help to
maintain the WIM \citep{haffner09}.  The WIM appears strongly
correlated in position and velocity with \hii\ regions
\citep{alves12}, although the particulars of the situation are not yet
well understood \citep{roshi12}.

In our ongoing HRDS census, we are trying to identify and observe all
possible Galactic \hii\ regions in RRL emission.  The detection of RRL
emission provides a velocity, which can then be turned into a
kinematic distance by assuming a Galactic rotation model
\citep[e.g.,][]{anderson09a, anderson12b}.  With a large sample of
\hii\ regions that have distances, we may better trace Galactic
structure and the properties of global massive star formation in the
Galaxy.  The HRDS has detected the RRL emission from over 800 nebulae
to date \citep{anderson11, bania12, anderson15c, brown17} spanning the
entire sky.
%In the first HRDS survey, we observed 448 previously
%unknown Galactic \hii\ regions over $67\degree > \ell > -17\degree$,
%$|b| \le 1\degree$, doubling the number known within the survey zone
%\citep{anderson11}.  In the Arecibo HRDS we discovered 37 more
%\hii\ regions \citep{bania12} over $67\degree > \ell > 32\degree$,
%$|b| \le 1\degree$, and we added 305 regions in the most recent HRDS
%survey covering the entire sky north of $\delta = -45\degree$
%($266\degree > \ell > -340\degree$ at $b = 0\degree$)w
%\citep{anderson14}.
For all HRDS surveys, targets were identified by matching mid-infrared
(MIR) emission [from {\it Spitzer} or the {\it Wide-field Infrared
    Survey Explorer (WISE)}] with public radio continuum data.
Previous HRDS surveys mostly targeted regions small with respect to
the telescope beam \citep[e.g.][]{anderson11}.

Here, we report Green Bank Telescope (GBT) observations of a sample of
angularly large \hii\ region candidates located North of $\delta =
-45\degree$ ($266\degree > \ell > -20\degree$ at $b = 0\degree$).
This survey completes the HRDS \hii\ region census in the Northern
sky.  Although they are $>1000$ \hii\ region candidates in the {\it
  WISE} Catalog remaining, these are too faint to be detected in RRL
emission for reasonable integration times.

\section{Target Selection}
We draw our targets from the {\it WISE} Catalog of Galactic
\hii\ Regions \citep{anderson14}, which lists $\sim2500$ \hii\ region
candidates identified from {\it WISE} data \citep{wright10} that have
associated radio continuum emission.  As in \citet{anderson15c}, we
also include in our sample Sharpless \hii\ regions not yet observed in
RRL emission. In \citet{anderson11}, we showed that the signature of
\hii\ regions in the MIR is $\sim10$\,\micron\ emission surrounding
$\sim20$\,\micron\ emission.  Our RRL detection rate of 95\% in
previous HRDS surveys shows that this criterion accurately identifies
{\it bona fide} \hii\ regions.  It can be applied using {\it Spitzer}
data (at 8.0\,\micron\ and 24\,\micron) as in \citet{anderson11,
  bania12}, and also {\it WISE} data (at 12\,\micron\ and 22\,\micron)
as in \citet{anderson15c}.

\begin{figure}
  \begin{centering}
  \includegraphics[width=3.25in]{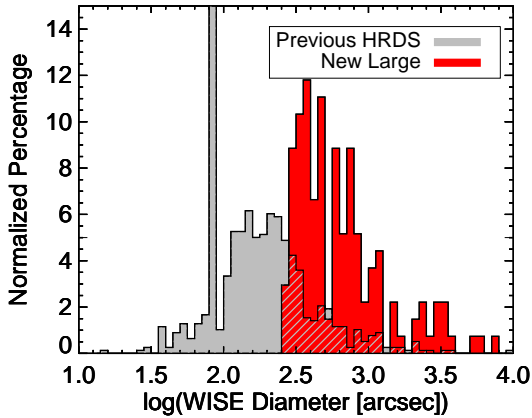}
  \caption{\hii\ region angular diameter distributions of detected new
    large (red) and previous HRDS (black) \hii\ regions.  All
      angular sizes come from the {\it WISE} Catalog of Galactic
      \hii\ regions \citep{anderson14}. The distributions are
    normalized such that the values in each bin show the percentage of
    the total distribution. Hatched areas show overlapping
    distributions.  The many regions with diameters near $85\arcsec$
    (or 1.8 along the x-axis) are an artifact due to how these
    sizes were defined in the {\it WISE} catalog; the true diameters for most such regions are slightly less than $85\arcsec$.  
    \label{fig:size}}
\end{centering}
\end{figure}

The candidates have angular diameters $>260\arcsec$ as defined from
MIR data in the {\it WISE} \hii\ region catalog.  Because the
  infrared size of an \hii\ region includes the emission from its
  PDR, the radio sizes of \hii\ regions are smaller than the infrared
sizes, on average by approximately a factor of two \citep{bihr16}.
Compared with that of previous HRDS surveys, our sample is skewed
toward angularly large regions (Figure~\ref{fig:size}).
%Large \hii\ regions are straightforward to identify from this
%criterion alone.  

%%%%%%%%%%%%%%%%%%%%%%%%%%%%%%%%%%%%%%%%%%%%%%%%%%%%%%%%%%%%%%%%%%%%%%%%%%%%%%%%
%%%%%%%%%%%%%%%%%%%%%%%%%%%%%%%%%%%%%%%%%%%%%%%%%%%%%%%%%%%%%%%%%%%%%%%%%%%%%%%%

All objects targeted here also have spatially coincident $\sim20$\,cm
continuum emission.  We use this 20\,cm continuum data to determine
the expected antenna temperature of our RRL observations.  We
primarily use the VLA Galactic Plane Survey 21\,cm continuum data
\citep[VGPS;][]{stil06} in the first Galactic quadrant, the Canadian
Galactic Plane Survey 21\,cm continuum data \citep[CGPS;][]{taylor03}
in the second Galactic quadrant, and 20\,cm NRAO VLA Sky Survey
otherwise \citep[NVSS;][]{condon98}.  We select only candidates that
have peak 20\,cm radio continuum emission of at least 45\,\mjy.  After
extrapolating the 20\cm emission to our observed 6\,cm wavelength
by assuming optically thin free-free continuum emission
($S_\nu\propto\nu^{-0.1}$), a RRL line-to-continuum flux density ratio
of 20, and a GBT gain of 2\,K\,Jy$^{-1}$, this results in a
lower-limit predicted RRL intensity of 5\,mK, or 2.5\,mJy.

\section{Observations and Data Reduction}
We observe our targets using the 100-m GBT at C-band (4--8\,\ghz).
Previous HRDS surveys used X-band (8--10\,\ghz) observations, but
these nebulae are better matched to the GBT C-band beam of $\sim
150\arcsec$.  Additionally, RRLs from low-density plasma are amplified
by stimulated emission, an effect that is frequency-dependent.  Plasma
of density $10^2\,\percc$, which is not unreasonable for many of these
regions, is amplified most strongly at C-band \citep{salem79}.  This
in principle makes the detection of RRL emission from large
\hii\ regions easier than it would be at other frequencies.  Finally,
hardware upgrades at the GBT make RRL observations with the GBT at
C-band extremely sensitive.

We employ total-power position-switching RRL spectroscopic
observations with six-minute on-source and off-source integrations,
hereafter a ``pair.''  The off-source positions are located $6\arcmin$
in Right Ascension from the on-source locations, which ensures that
they track the same path on the sky.  We observe nearly all sources
for two pairs, or 12-minutes on-source.  We alter this integration
time slightly such that the integration times for the brightest
extrapolated C-band continuum emission are 6-minutes (one pair), and
the faintest one 24-minutes (four pairs).  Unlike in previous GBT HRDS
surveys, due to the low surface brightness of these targets and
confusion within the Galactic plane, we do not measure their continuum
emission.

The GBT recently upgraded both its spectrometer and also the C-band
receiver.  The new spectrometer, called the Versatile GBT Astronomical
Spectrometer \citep[VEGAS;][]{prestage15}, can simultaneously observe
64 spectral windows.  The upgraded C-band receiver has a
3.80\,\ghz\ instantaneous bandwidth, from 3.95 to 8.00\,\ghz. Using
the VEGAS/C-band combination, we simultaneously observe
22~\hna\ transitions from H117$\alpha$ to H95$\alpha$, not including
H113$\alpha$ (which is compromised by H142$\beta$).  These transitions
span frequencies from 4.05 to 7.55\,\ghz.  We observe each line in two
orthogonal linear polarizations.

We made extensive tests of the VEGAS/C-band configuration to
ensure stability across the bandpass, and to verify the intensity
calibration.  Because of instabilities at the extremes of the $\sim
4$\,\ghz\ instantaneous bandpass, we disregard data from the
H117$\alpha$, H116$\alpha$, H115$\alpha$, H96$\alpha$, and H95$\alpha$
lines in our analysis.  The total usable instantaneous bandpass is
$\sim 4.3\,\ghz$ to $\sim 7.2\,\ghz$, which includes 17 uncompromised
hydrogen RRLs.  We verified the performance of our configuration by
observing the strong RRL source W3 at the beginning of each observing
session. The measured hydrogen RRL intensities for W3 across the
usable bandpass agree with those measured in previous work
\citep{balser16}.  Furthermore, the intensities of the W3 RRLs were
stable in time, and invariant with weather, to within 10\%.

We initially calibrated the spectra using the measured temperature of
a noise diode whose power was periodically injected into the signal
path during observations.  In previous work with the GBT at X-band, we
found that this calibrates the data to within 10\%
\citep[e.g.,][]{anderson15c}.  We additionally observed the primary
flux calibrator 3C286 in two modes: 1) by making small fully-sampled
maps scanning in R.A. and Dec, and 2) using total power
position-switched observations.  For both modes, we use the same
spectrometer configuration as for the science targets.  In previous
HRDS surveys, we used the Digital Continuum Receiver (DCR) to perform
flux calibration measurements by tuning the DCR to each observed
frequency.  Because of the large number of frequencies we now observe
simultaneously, it is not feasible to use the DCR by tuning to each
frequency individually.  We derive the continuum intensities of 3C286
from the vertical offset of the spectral baselines, i.e., the change
in system temperature as we switch from on- to off-source positions,
and assume a gain of 2\,K\,Jy$^{-1}$.  For the maps, we find that the
derived flux densities averaged over all \hna\ spectral windows agree
with the values for 3C286 given in \citet{ott94} to within 4\%, for
all weather conditions.  For the pointed observations, the agreement
between the noise-diode calibrated data and the data for 3C286 is good
to within $\sim 10\%$ in good weather, and $\sim 15\%$ in poorer
weather.  Presumably the discrepancy between the results of the two
modes is due to better sky subtraction in the maps.  The C-band
intensity scale is relatively insensitive to opacity and elevation
gain corrections, which both have magnitudes of $\lsim 5\%$
\citep{ghigo01}. We conclude that the absolute intensities of the
derived line parameters are good to within 15\%, even in the worst
observing conditions.  We make no attempt to correct for calibration
uncertainties, opacity effects, or elevation effects.

We follow the same data reduction steps as in previous HRDS surveys,
employing the TMBIDL software package \citep[][V8, Zenodo, doi:10.5281/zenodo.32790]{bania16}.  We
average the two polarizations of each transition, then average all 17
usable \hna\ transitions after regridding to the velocity scale of the
H95$\alpha$ transition.  This line-stacking method was originally
demonstrated by \citet{balser06}, and later employed in all HRDS
surveys.  Averaging the spectra in this way improves the RRL
signal-to-noise ratio, allowing for significantly reduced integration
times.  We remove a low-order polynomial baseline (typically 4th
order), smooth to 1.82\,\kms\ resolution, and fit Gaussian line
profiles to the resultant spectrum.  Thus, we derive the
LSR\footnote{Defined as the kinematic local standard of rest (LSR)
  frame using the radio Doppler shift convention.  The kinematic LSR
  is defined by a solar motion of 20.0\,\kms\ toward $(\alpha, \delta)
  = (18^{\rm h}, +30\degree) [1900.0]$ \citep{gordon76}.}  velocity,
line intensity, and full width at half-maximum (FWHM) line width for
each hydrogen RRL component.  We identify the highest-velocity line as
being from hydrogen.  We also identify helium and carbon RRLs, offset
by $-122.15$ and $-149.56\,\kms$ from the hydrogen RRLs, respectively.

The GBT beam at the observed frequencies ranges from $170\arcsec$
($2.9\arcmin$) to $110\arcsec$ ($1.77\arcmin$), a difference of 60\%.
The regions observed all have {\it WISE}-derived diameters of
$260\arcsec$ or greater \citep[and therefore radio-diameters of
  perhaps $\sim 130\arcsec$, c.f.][]{bihr16}.  Because many of these
regions are resolved, each frequency may sample a slightly different
portion of each region.  We attempt no correction for this
complication.

\section{The Catalog of Large H\,{\bf\footnotesize II} Regions\label{sec:catalog}}
\begin{figure*}
  \begin{centering}
    \includegraphics[width=3.5in]{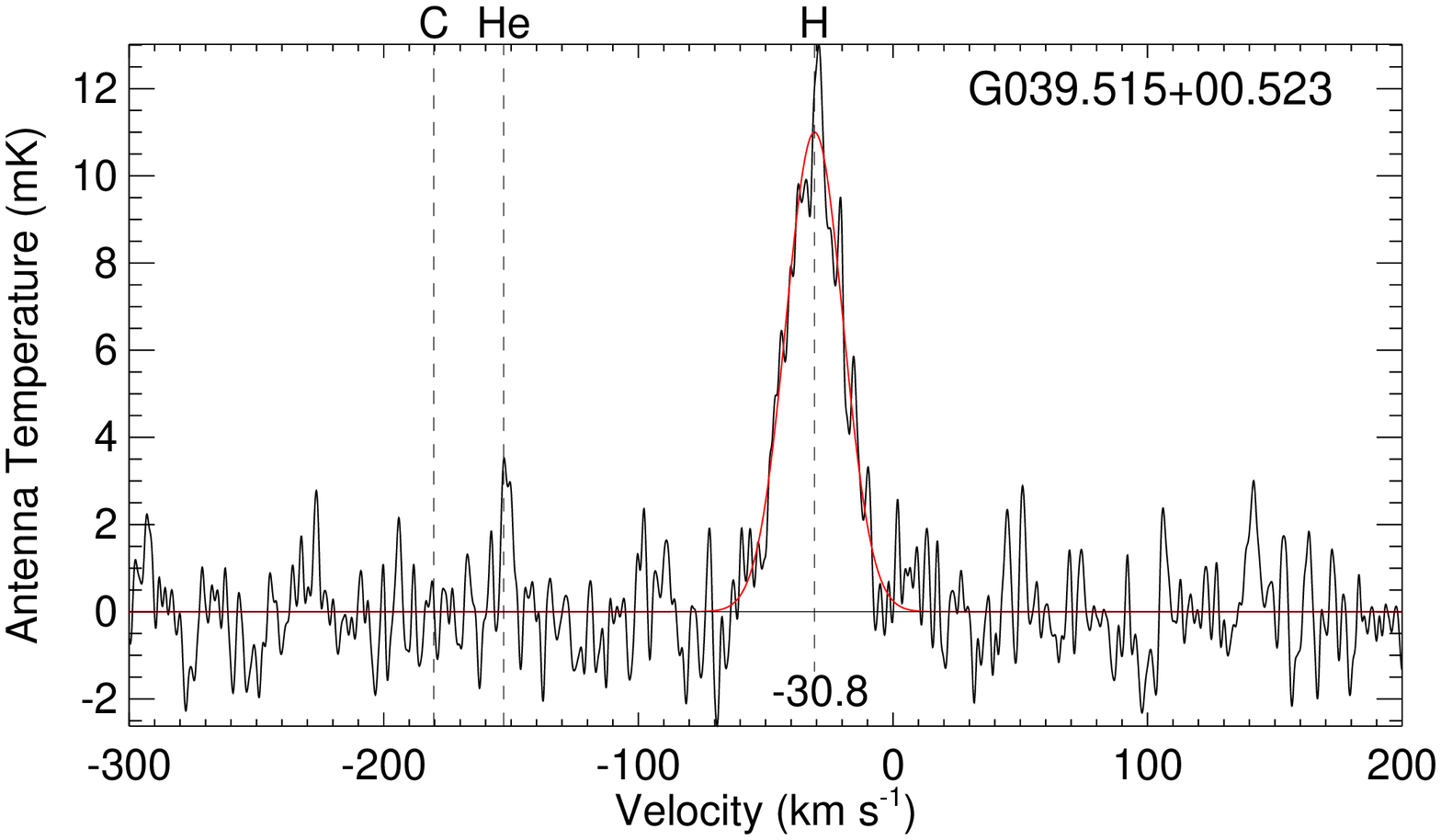}
    \includegraphics[width=2 in]{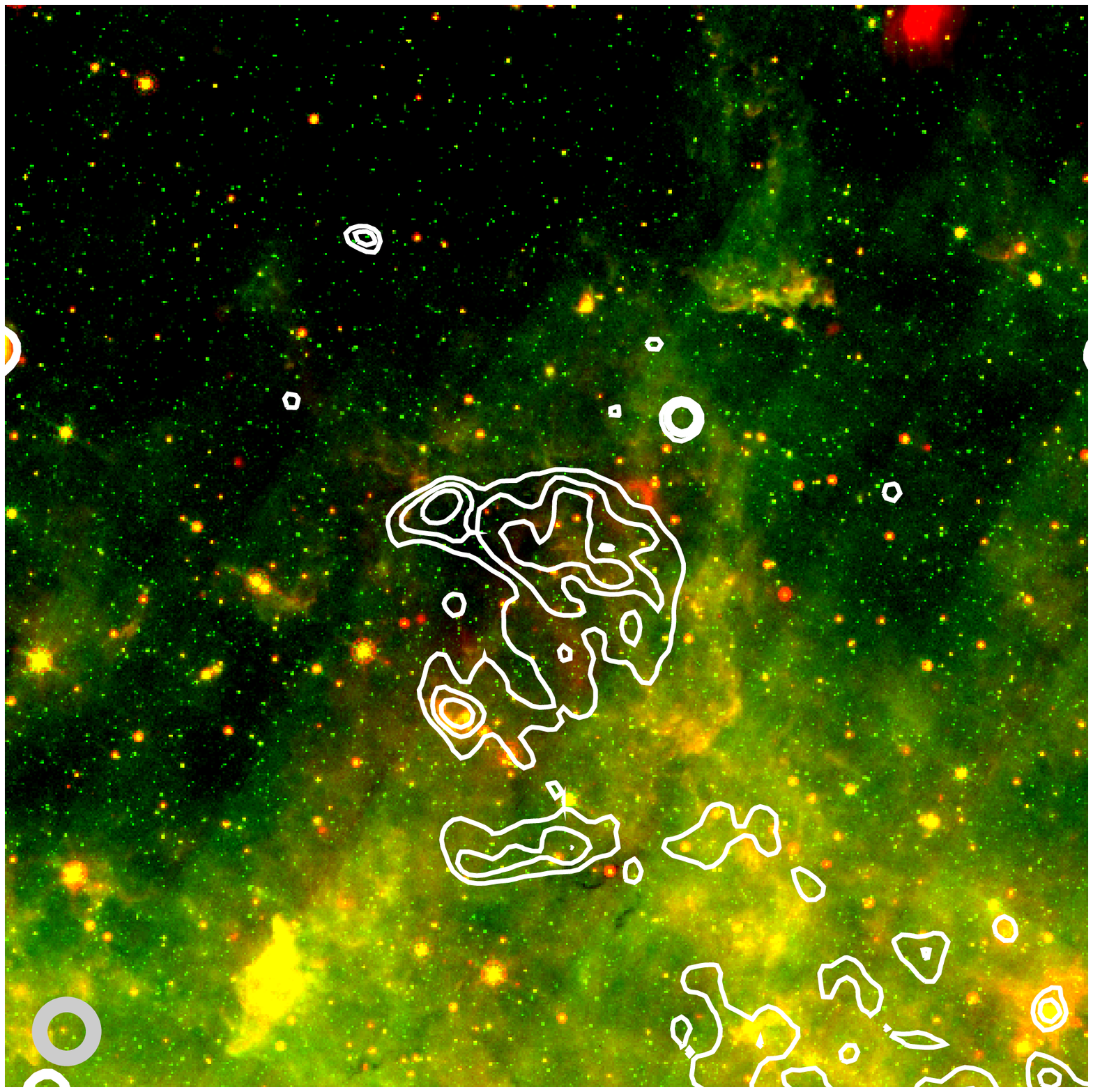}
    \includegraphics[width=3.5in]{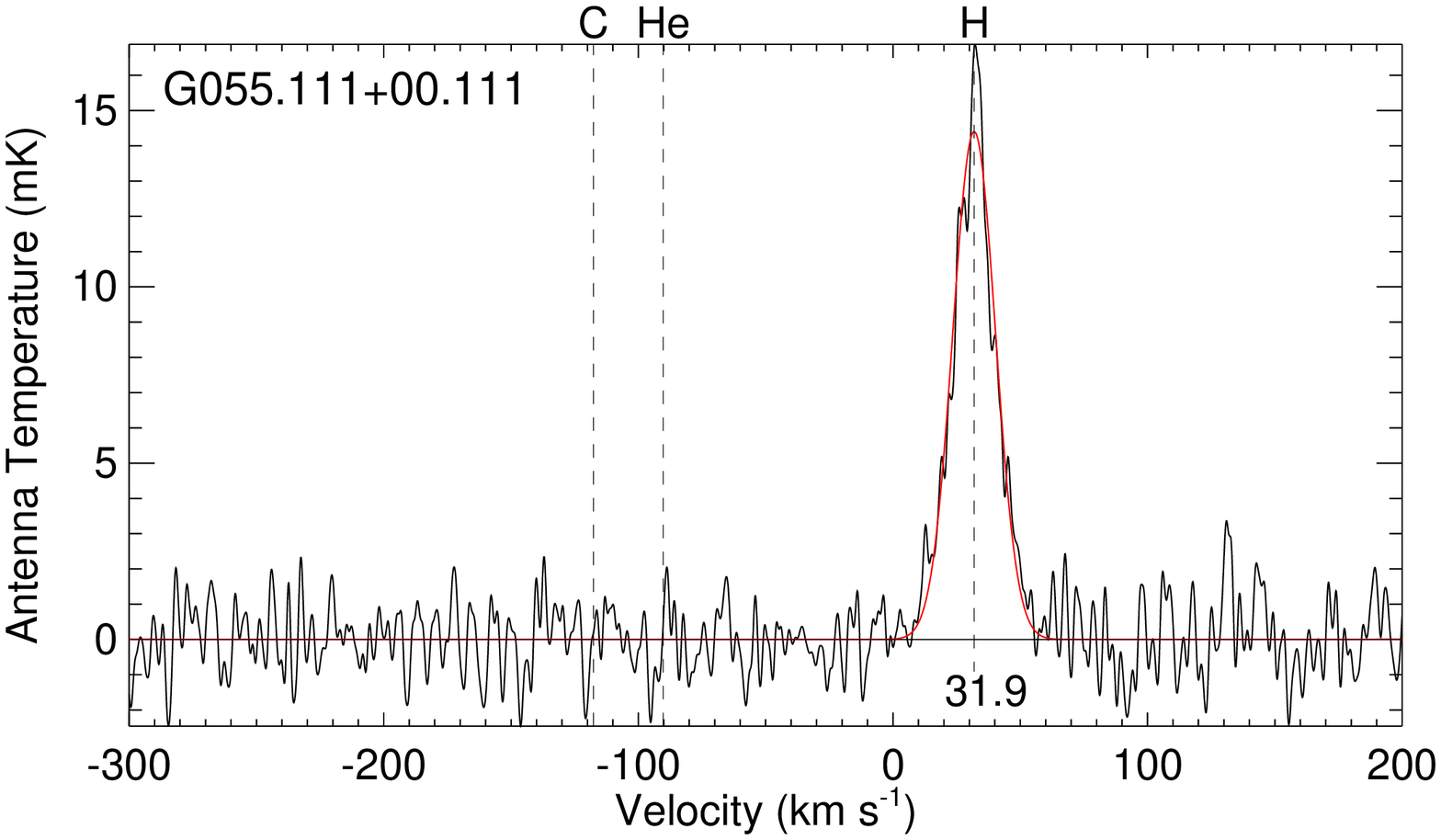}
    \includegraphics[width=2 in]{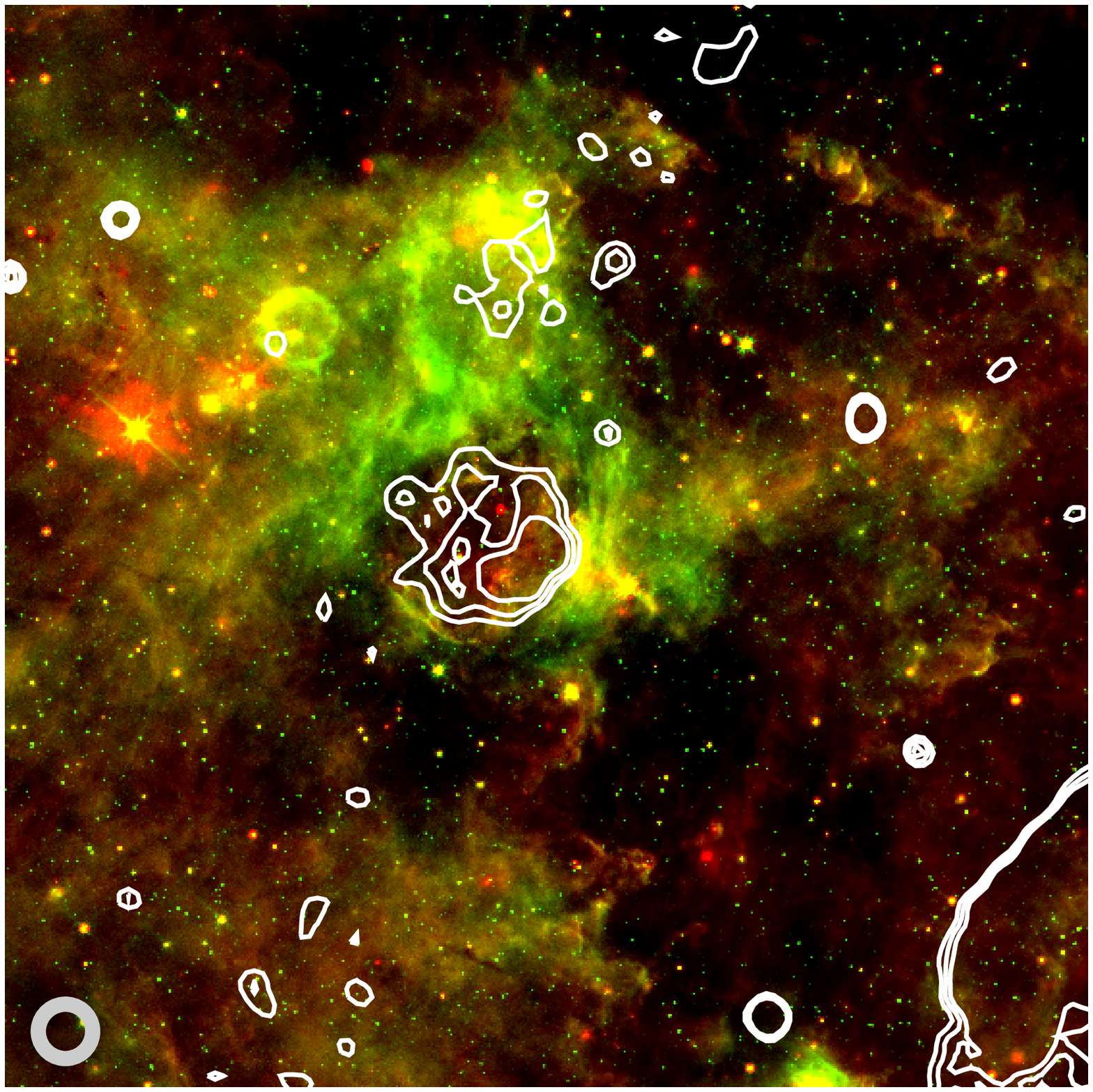}
    \includegraphics[width=3.5in]{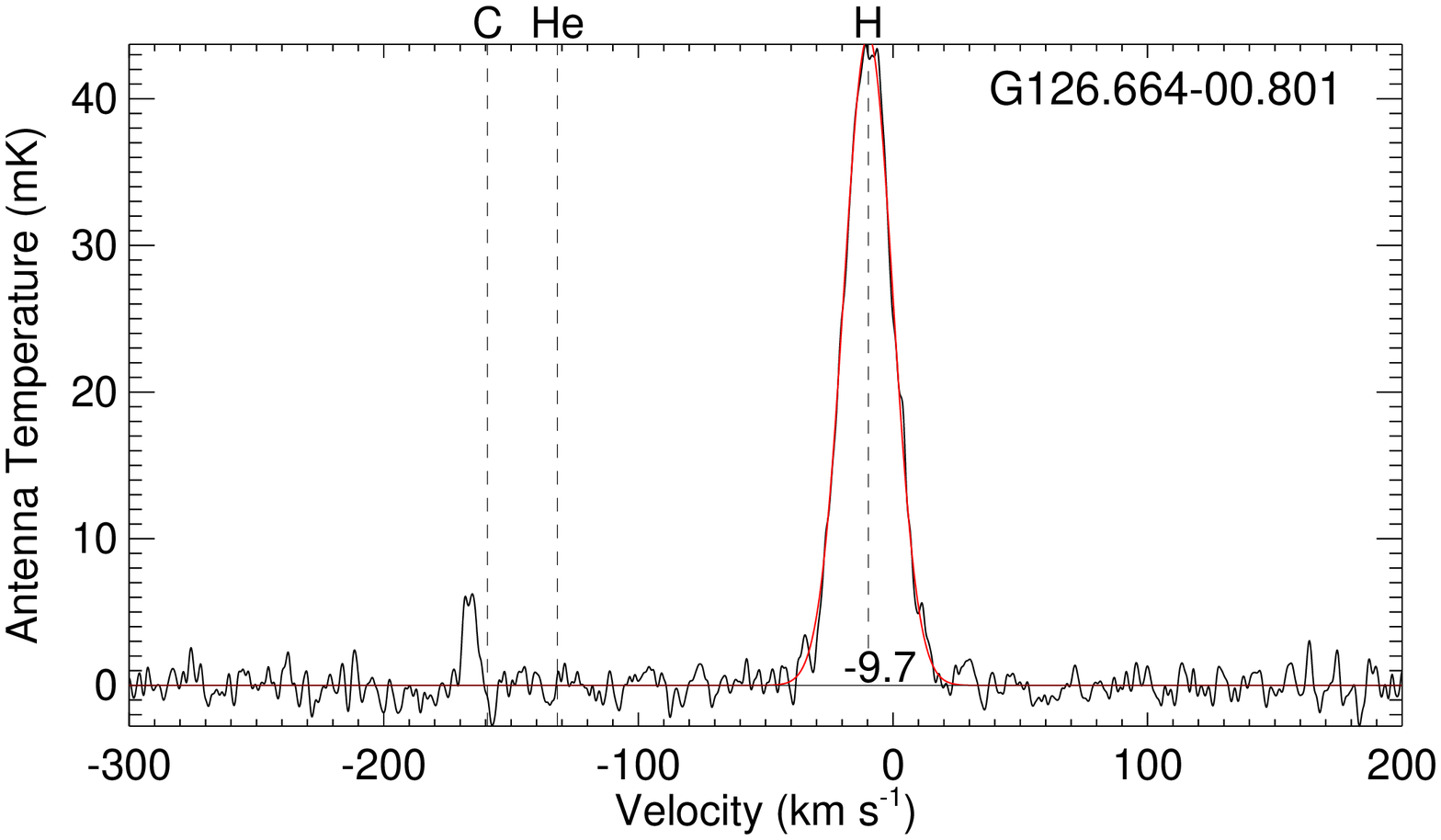}
    \includegraphics[width=2 in]{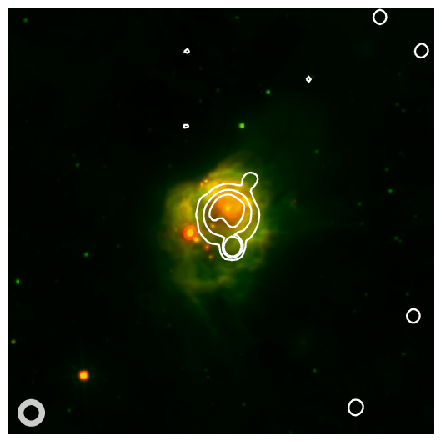}
    \caption{Representative spectra and images of observed large
      \hii\ regions G039.515+00.523 (top), G055.111+00.011 (middle),
      and S187 (bottom).  The spectra are averages of RRL transitions
      H114$\alpha$ to H97$\alpha$, smoothed to 1.82\,\kms, with
      hydrogen RRL fits shown in red.  The expected velocities of the
      helium and carbon RRLs are also indicated.  S187 has a weak
      carbon line detected.  Images are $40\arcmin$ square centered at
      the observed location and show {\it Spitzer} MIPSGAL
      24\,\micron\ (red) and GLIMPSE 8.0\,\micron\ (green) for
      G039.515+00.523 and G055.111+00.011 or {\it WISE}
      22\,\micron\ (red) and {\it WISE} 12\,\micron\ (green) for S187.
      White contours are of 21\,cm continuum emission, showing the
      location of the ionized gas.  For G039.515+00.523 and
      G055.111+00.011 the data come from the VGPS and have levels of
      14,15, and 16\,\K and 9, 9.5, and 10\,\K, respectively.  For
      S187 the data come from the CGPS and have levels of 6, 10, and
      15\,\K.  The $150\arcsec$ GBT beam is shown in thick gray
      circles at the lower left of each image.\label{fig:example}}
\end{centering}
\end{figure*}

We detect hydrogen RRLs from 148 of the 157 observed sources, for a
94\% detection rate.  This is comparable to the 95\% detection rate of
the original GBT HRDS \citep{anderson11} and the 93\% detection rate
of the GBT HRDS extension \citet{anderson15c}.  Twenty-three of the
148 detections are of Sharpless \hii\ regions not previously detected
in RRL emission.  We give the derived hydrogen RRL parameters in
Table~\ref{tab:line}, which lists the source name, the Galactic
longitude and latitude, the line intensity, the FWHM line width, the
LSR velocity, and the r.m.s.  noise in the spectrum.  The errors given
in Table \ref{tab:line} for the line parameters are the $1\,\sigma$
uncertainties from the Gaussian fits.  For sources with multiple
velocity components detected along the line of sight, we append to
their names additional letters ``a'', ``b'', or ``c'' in order of
decreasing peak line intensity.  All listed hydrogen lines have a
signal-to-noise ratio (SNR) of at least 5, where the SNR as defined by
\citet{lenz92} is
\begin{equation}
{\rm SNR} = 0.7 \left( \frac{T_L}{\rm r.m.s.}\right) \left( \frac{\Delta V}{1.82} \right)^{0.5}\,,
\label{eq:snr}
\end{equation}
%For multiple-velocity sources, if we were able to determine which
%component stems from the discrete \hii\ region (see
%Section~\ref{sec:multvel}) it is flagged with an asterisk in the final
%column of Table~\ref{tab:line}.
where $T_L$ is the peak line height, r.m.s. is the root-mean-squared
spectral noise, $\Delta V$ is FWHM line width in \kms, and the factor of 1.82
is the FWHM of the Gaussian smoothing kernel we used.  We show example
spectra in Figure~\ref{fig:example}.

%%%%%%%%%%%%%%%%%%%%%%%%%%%%%%%%%%%%%%%%%%%%%%%%%%%%%%%%%%%%%%%%%%%%%%%
\begin{deluxetable*}{lccrcccrccc}
\tabletypesize{\scriptsize}
%\rotate
\tablecaption{Hydrogen Recombination Line Parameters}
\tablewidth{0pt}
\tablehead{
\colhead{Source\tablenotemark{a}} &
\colhead{\gl} &
\colhead{\gb} &
\colhead{$T_L$} &
\colhead{$\sigma T_L$} &
\colhead{$\Delta V$} &
\colhead{$\sigma \Delta V$} &
\colhead{$V_{LSR}$} &
\colhead{$\sigma V_{LSR}$} &
\colhead{r.m.s.}%&
%\colhead{Note\tablenotemark{a}}
\\
\colhead{} &
\colhead{($\arcdeg$)} &
\colhead{($\arcdeg$)} &
\colhead{(mK)} &
\colhead{(mK)} &
\colhead{(\kms)} &
\colhead{(\kms)} &
\colhead{(\kms)} &
\colhead{(\kms)} &
\colhead{(mK)}% &
%\colhead{}
}
\startdata
S17a &    0.003 & +0.128 & $\phn 94.8$  &  $0.9$  &  $13.6$  &  $0.2$  &  \phn \phn 7.5  &  $0.1$ & 1.7 \\
S17b &    0.003 & +0.128 & $\phn 89.9$  &  $0.6$  &  $20.1$  &  $0.3$  &  \phn $-12.1$  &  $0.1$ & 1.7 \\
S19 &    0.124 & $-$0.568 & $\phn 92.4$  &  $0.5$  &  $17.1$  &  $0.1$  &  \phn 15.5  &  $0.1$ & 2.3 \\
G003.253$-$00.280 &    3.253 & $-$0.280 & $\phn \phn 7.2$  &  $0.1$  &  $25.7$  &  $0.4$  &  \phn \phn 5.1  &  $0.2$ & 1.0 \\
G003.597$-$00.206a &    3.597 & $-$0.206 & $\phn 13.3$  &  $0.1$  &  $15.6$  &  $0.2$  &  \phn $-27.0$  &  $0.1$ & 1.0 \\
G003.597$-$00.206b &    3.597 & $-$0.206 & $\phn \phn 3.5$  &  $0.1$  &  $36.2$  &  $1.2$  &  \phn \phn 8.8  &  $0.5$ & 1.0 \\
G003.713$-$00.215a &    3.713 & $-$0.215 & $\phn \phn 8.9$  &  $0.1$  &  $23.7$  &  $0.3$  &  \phn \phn 4.1  &  $0.1$ & 1.0 \\
G003.713$-$00.215b &    3.713 & $-$0.215 & $\phn \phn 4.5$  &  $0.1$  &  $15.1$  &  $0.5$  &  \phn $-28.9$  &  $0.2$ & 1.0 \\
G006.389$-$00.198 &    6.389 & $-$0.198 & $\phn 75.7$  &  $0.3$  &  $19.0$  &  $0.1$  &  \phn 23.7  &  $0.1$ & 1.3 \\
S31 &    7.205 & $-$2.188 & $\phn 10.1$  &  $0.2$  &  $17.7$  &  $0.4$  &  \phn \phn 4.2  &  $0.2$ & 1.1 \\
G007.221$-$00.212 &    7.221 & $-$0.212 & $\phn 52.1$  &  $0.2$  &  $26.3$  &  $0.1$  &  \phn 25.6  &  $0.1$ & 1.0 \\
S32 &    7.294 & $-$2.044 & $\phn 24.9$  &  $0.2$  &  $18.1$  &  $0.2$  &  \phn \phn 6.0  &  $0.1$ & 1.1 \\
G007.380$-$00.264a &    7.380 & $-$0.264 & $\phn 49.5$  &  $0.2$  &  $14.2$  &  $0.1$  &  \phn 31.8  &  $0.1$ & 1.1 \\
G007.380$-$00.264b &    7.380 & $-$0.264 & $\phn 10.3$  &  $0.2$  &  $23.3$  &  $0.8$  &  \phn \phn 9.7  &  $0.3$ & 1.1 \\
G007.485$-$00.188a &    7.485 & $-$0.188 & $\phn 62.0$  &  $0.3$  &  $11.9$  &  $0.1$  &  \phn 48.4  &  $0.1$ & 1.1 \\
G007.485$-$00.188b &    7.485 & $-$0.188 & $\phn 12.3$  &  $0.2$  &  $21.1$  &  $0.6$  &  \phn \phn 3.7  &  $0.2$ & 1.1 \\
G007.485$-$00.188c &    7.485 & $-$0.188 & $\phn 12.0$  &  $0.2$  &  $17.9$  &  $0.9$  &  \phn 31.5  &  $0.3$ & 1.1 \\
\enddata
\label{tab:line}
\tablenotetext{a}{Source names for HII regions with multiple detected hydrogen RRL components are appended by ``a,'' ``b,'' or ``c,'' in order of decreasing hydrogen RRL intensity.}
\tablecomments{Table~\ref{tab:line} is available in its entirety in the electronic edition of the {\it Astrophysical Journal}. 
A portion is shown here for guidance regarding its form and content.}
\end{deluxetable*}

%%%%%%%%%%%%%%%%%%%%%%%%%%%%%%%%%%%%%%%%%%%%%%%%%%%%%%%%%%%%%%%%%%%%%%%%%%%%%%%%

%%%%%%%%%%%%%%%%%%%%%%%%%%%%%%%%%%%%%%%%%%%%%%%%%%%%%%%%%%%%%%%%%%%%%%%%%%%%%%%%

%%%%%%%%%%%%%%%%%%%%%%%%%%%%%%%%%%%%%%%%%%%%%%%%%%%%%%%%%%%%%%%%%%%%%%%
\begin{deluxetable*}{lccrcccrccc}
\tabletypesize{\scriptsize}
%\rotate
\tablecaption{Helium Recombination Line Parameters}
\tablewidth{0pt}
\tablehead{
\colhead{Source\tablenotemark{a}} &
\colhead{\gl} &
\colhead{\gb} &
\colhead{$T_L$} &
\colhead{$\sigma T_L$} &
\colhead{$\Delta V$} &
\colhead{$\sigma \Delta V$} &
\colhead{$V_{LSR}$} &
\colhead{$\sigma V_{LSR}$} &
\colhead{r.m.s.}%&
%\colhead{Note\tablenotemark{a}}
\\
\colhead{} &
\colhead{($\arcdeg$)} &
\colhead{($\arcdeg$)} &
\colhead{(mK)} &
\colhead{(mK)} &
\colhead{(\kms)} &
\colhead{(\kms)} &
\colhead{(\kms)} &
\colhead{(\kms)} &
\colhead{(mK)}% &
%\colhead{}
}
\startdata
S17a &    0.003 & +0.128 & $\phn 6.4$  &  $0.3$  &  $\phn 6.3$  &  $0.3$  &  \phn \phn 6.8  &  $0.1$ & 1.7 \\
S19 &    0.124 & $-$0.568 & $\phn 9.8$  &  $0.3$  &  $12.0$  &  $0.1$  &  \phn 15.6  &  $0.2$ & 2.3 \\
G007.221$-$00.212 &    7.221 & $-$0.212 & $\phn 4.1$  &  $0.1$  &  $19.0$  &  $0.4$  &  \phn 24.2  &  $0.2$ & 1.0 \\
G007.485$-$00.188a &    7.485 & $-$0.188 & $\phn 7.0$  &  $0.2$  &  $\phn 3.6$  &  $0.2$  &  \phn 49.1  &  $0.1$ & 1.1 \\
G007.700$-$00.122a &    7.700 & $-$0.122 & $\phn 6.7$  &  $0.2$  &  $10.3$  &  $0.4$  &  \phn \phn $-3.9$  &  $0.2$ & 1.2 \\
G018.594+00.320 &   18.594 & +0.320 & $\phn 4.9$  &  $0.1$  &  $14.6$  &  $0.3$  &  \phn 12.0  &  $0.1$ & 1.2 \\
G019.452+00.117a &   19.452 & +0.117 & $12.0$  &  $0.2$  &  $\phn 9.5$  &  $0.2$  &  \phn 20.3  &  $0.1$ & 2.0 \\
G031.006+00.186a &   31.006 & +0.186 & $\phn 4.5$  &  $0.1$  &  $12.4$  &  $0.3$  &  107.2  &  $0.1$ & 1.0 \\
G035.335+00.007 &   35.335 & +0.007 & $\phn 3.4$  &  $0.1$  &  $16.0$  &  $0.5$  &  \phn 57.2  &  $0.2$ & 1.1 \\
G035.623$-$00.217 &   35.623 & $-$0.217 & $\phn 4.1$  &  $0.2$  &  $12.8$  &  $0.6$  &  \phn 48.4  &  $0.2$ & 1.4 \\
G037.838$-$00.509 &   37.838 & $-$0.509 & $\phn 3.5$  &  $0.1$  &  $14.4$  &  $0.6$  &  \phn 62.2  &  $0.2$ & 1.2 \\
G039.299$-$00.380 &   39.299 & $-$0.380 & $\phn 2.5$  &  $0.1$  &  $18.9$  &  $0.9$  &  \phn 52.4  &  $0.4$ & 0.9 \\
G076.156+00.636 &   76.156 & +0.636 & $\phn 5.1$  &  $0.1$  &  $17.2$  &  $0.5$  &  \phn 11.1  &  $0.2$ & 1.2 \\
G213.077$-$02.217 &  213.077 & $-$2.217 & $\phn 4.4$  &  $0.1$  &  $12.1$  &  $0.4$  &  \phn 58.7  &  $0.2$ & 1.3 \\
G340.211$-$00.318 &  340.211 & $-$0.318 & $\phn 8.7$  &  $0.2$  &  $11.2$  &  $0.4$  &  \phn $-46.5$  &  $0.2$ & 2.2 \\
G341.668$-$00.398 &  341.668 & $-$0.398 & $\phn 6.2$  &  $0.2$  &  $14.7$  &  $0.6$  &  \phn $-29.7$  &  $0.3$ & 2.1 \\
G342.120+00.002a &  342.120 & +0.002 & $10.3$  &  $0.2$  &  $18.6$  &  $0.3$  &  $-131.1$  &  $0.1$ & 2.1 \\
G344.472$-$00.556 &  344.472 & $-$0.556 & $\phn 5.2$  &  $0.2$  &  $\phn 9.2$  &  $0.4$  &  \phn $-20.4$  &  $0.2$ & 1.6 \\
G344.753$-$00.547 &  344.753 & $-$0.547 & $\phn 4.4$  &  $0.2$  &  $\phn 9.8$  &  $0.5$  &  \phn $-26.4$  &  $0.2$ & 1.3 \\
G346.261+00.313 &  346.261 & +0.313 & $\phn 3.3$  &  $0.1$  &  $19.4$  &  $0.7$  &  \phn \phn 5.1  &  $0.3$ & 1.2 \\
G355.117+00.014a &  355.117 & +0.014 & $\phn 3.9$  &  $0.2$  &  $11.7$  &  $0.7$  &  \phn 14.1  &  $0.3$ & 1.1 \\

\enddata
\label{tab:he_line}
\tablenotetext{a}{Source names for HII regions with multiple detected hydrogen RRL components are appended by ``a'' or ``b,'' in order of decreasing hydrogen RRL intensity.}
\end{deluxetable*}
%%%%%%%%%%%%%%%%%%%%%%%%%%%%%%%%%%%%%%%%%%%%%%%%%%%%%%%%%%%%%%%%%%%%%%%%%%%%%%%%

%%%%%%%%%%%%%%%%%%%%%%%%%%%%%%%%%%%%%%%%%%%%%%%%%%%%%%%%%%%%%%%%%%%%%%%
\begin{deluxetable*}{lccrcccrccc}
\tabletypesize{\scriptsize}
%\rotate
\tablecaption{Carbon Recombination Line Parameters}
\tablewidth{0pt}
\tablehead{
\colhead{Source} &
\colhead{\gl} &
\colhead{\gb} &
\colhead{$T_L$} &
\colhead{$\sigma T_L$} &
\colhead{$\Delta V$} &
\colhead{$\sigma \Delta V$} &
\colhead{$V_{LSR}$} &
\colhead{$\sigma V_{LSR}$} &
\colhead{r.m.s.}%&
%\colhead{Note\tablenotemark{a}}
\\
\colhead{} &
\colhead{($\arcdeg$)} &
\colhead{($\arcdeg$)} &
\colhead{(mK)} &
\colhead{(mK)} &
\colhead{(\kms)} &
\colhead{(\kms)} &
\colhead{(\kms)} &
\colhead{(\kms)} &
\colhead{(mK)}% &
%\colhead{}
}
\startdata
G007.221$-$00.212 &    7.221 & $-$0.212 & $\phn 3.6$  &  $0.2$  &  $3.1$  &  $0.2$  &  26.5  &  $0.1$ & 1.0 \\
S32 &    7.294 & $-$2.044 & $\phn 5.8$  &  $0.1$  &  $2.8$  &  $0.1$  &  10.4  &  $0.1$ & 1.1 \\
G058.607+00.619 &   58.607 & +0.619 & $\phn 3.4$  &  $0.1$  &  $5.2$  &  $0.2$  &  30.3  &  $0.1$ & 1.0 \\
G063.127+00.347 &   63.127 & +0.347 & $\phn 2.8$  &  $0.2$  &  $5.1$  &  $0.4$  &  18.1  &  $0.2$ & 1.0 \\
S149 &  108.375 & $-$1.056 & $17.3$  &  $0.3$  &  $4.6$  &  $0.1$  &  $-53.5$  &  $0.1$ & 1.1 \\
G108.907+02.693 &  108.907 & +2.693 & \phn$8.2$  &  $0.1$  &  $2.3$  &  $0.1$  &  $-10.7$  &  $0.1$ & 0.8 \\
G122.774+02.855 &  122.774 & +2.855 & \phn$8.1$  &  $0.2$  &  $2.5$  &  $0.1$  &  $-63.7$  &  $0.1$ & 0.7 \\
S187 &  126.664 & $-$0.801 & $\phn 6.9$  &  $0.8$  &  $5.6$  &  $0.8$  &  $-16.7$  &  $0.3$ & 0.9 \\
G188.546+03.668 &  188.546 & +3.668 & $27.1$  &  $0.1$  &  $2.6$  &  $0.1$  &  \phn $-6.7$  &  $0.1$ & 0.9 \\
G189.768+00.335 &  189.768 & +0.335 & $38.6$  &  $0.2$  &  $3.1$  &  $0.1$  &  \phn 8.7  &  $0.1$ & 0.9 \\
G190.051+00.537 &  190.051 & +0.537 & $25.0$  &  $0.1$  &  $3.3$  &  $0.1$  &  \phn 8.3  &  $0.1$ & 1.2 \\
S273 &  202.997 & +2.062 & $\phn 7.3$  &  $0.1$  &  $3.4$  &  $0.1$  &  10.2  &  $0.1$ & 1.0 \\
G207.326$-$02.140 &  207.326 & $-$2.140 & $10.7$  &  $0.2$  &  $3.5$  &  $0.1$  &  15.4  &  $0.1$ & 1.3 \\
G213.077$-$02.217 &  213.077 & $-$2.217 & $13.3$  &  $0.1$  &  $4.8$  &  $0.1$  &  50.7  &  $0.1$ & 1.3 \\
G221.852$-$02.024 &  221.852 & $-$2.024 & $\phn 6.5$  &  $0.1$  &  $4.2$  &  $0.1$  &  39.5  &  $0.1$ & 1.6 \\
S296 &  224.377 & $-$2.082 & $\phn 5.2$  &  $0.1$  &  $3.5$  &  $0.1$  &  17.5  &  $0.1$ & 1.3 \\

\enddata
\label{tab:c_line}
\end{deluxetable*}
%%%%%%%%%%%%%%%%%%%%%%%%%%%%%%%%%%%%%%%%%%%%%%%%%%%%%%%%%%%%%%%%%%%%%%%%%%%%%%%%

We additionally fit the helium and carbon RRLs.  We detect 21 helium
lines (Table~\ref{tab:he_line}) with a SNR as defined in
Equation~\ref{eq:snr} of at least 5 and 16 carbon lines
(Table~3) with a SNR of at least 3 (the narrow width of
the carbon lines allows for a lower SNR threshold since it is less
likely to be confused with background fluctuations).  These data will
be analyzed in a future publication; we provide them here for
completeness.  Although most of the carbon lines are weak, that they
are found at the expected offset velocities gives us additional
confidence in their detections.  Due to its higher atomic mass, the
helium line widths should be less than that of hydrogen, perhaps by
25\% to 50\% (depending on the amount of turbulence).  The smoothing
of 1.82\,\kms\ therefore has minimal impact on the derived helium line
parameters.  The carbon lines, however, are much narrower than those
of hydrogen, due to carbon's higher atomic mass and because carbon RRL
originate in cooler \hii\ region photo-dissociation regions.
%% The ``true'' carbon line
%% widths can be estimated by
%% \begin{equation}
%%   \Delta V_{\rm true} = \left( \Delta V_{\rm meas}^2 + 1.82^2 \right)^{0.5}\,\kms\,,
%% \end{equation}
%% where $\Delta V_{\rm true}$ and $\Delta V_{\rm meas}$ are the ``true''
%% and measured carbon RRL line widths, and all parameters are given in
%% \kms.  The ``true'' carbon peak line intensities can then be estimated
%% by keeping the integrated line intensity the same, i.e.,
%% \begin{equation}
%%   T_{L, {\rm true}} = T_{L, {\rm meas}} \left( \frac{\Delta V_{\rm
%%       meas}}{\Delta V_{\rm true}} \right)\,.
%% \end{equation}

\begin{figure}
  \begin{centering}
    \includegraphics[width=3.25in]{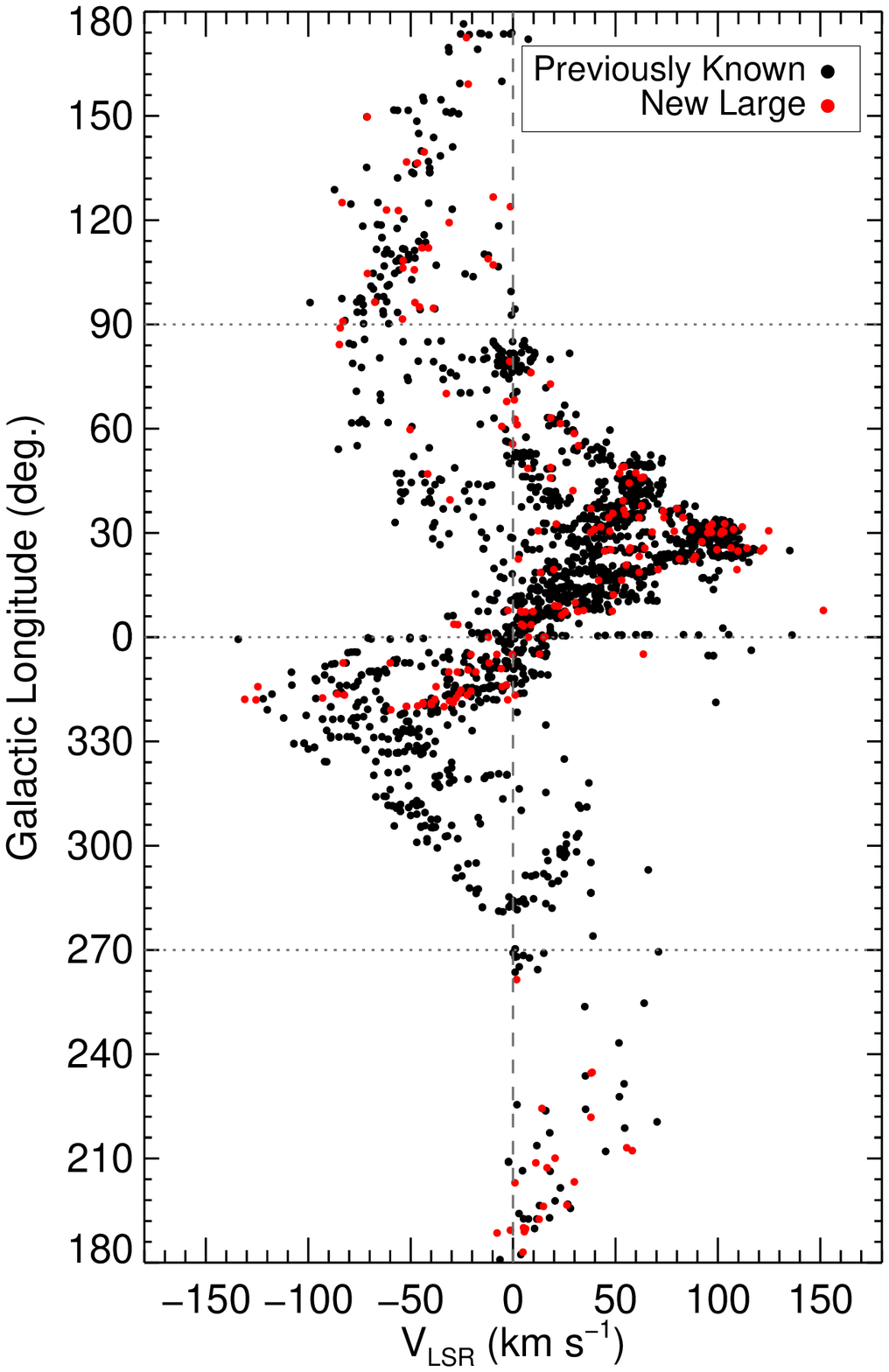}
  \caption{Longitude-velocity diagram of new large (red) and
    previously-known (black) \hii\ regions, the latter compiled in the
    {\it WISE} Catalog.  The two samples share similar distributions,
    although there is a higher proportion of new large regions in the
    third Galactic quadrant, and a lower proportion in the outer first
    Galactic quadrant (at negative velocities).\label{fig:lv}}
  \end{centering}
\end{figure}

The new large regions appear to have a similar Galactic distribution
compared with the previously-known sample of Galactic \hii\ regions.
This is also true of the ``diffuse'' region sample reported by
\citet{lockman96}.  In Figure~\ref{fig:lv} we show the
longitude-velocity location of the new large and previously-known
regions compiled in the {\it WISE} Catalog.  We note two slight
discrepancies between the distributions.  First, there are 21 new
third-quadrant detections, versus 36 previously-known, a 37\% increase
in the known \hii\ region population.  Over the rest of the present
survey, the 128 detections in the first, second, and fourth Galactic
quadrants only represent a 9\% increase in the known \hii\ region
population.  %{\bf This third-quadrant increase could be explained if
  %the sources are on average near to the Sun, they are angularly large
%and included in our sample.
Second, we only detect 10 negative-velocity \hii\ regions (7\%) in the
first Galactic quadrant (restricted to $\ell < 70\degree$), whereas
the combination of results from previous HRDS surveys gives 13\%.
Such regions have large Heliocentric distances and are located outside
of the Solar orbit.

%%%%%%%%%%%%%%%%%%%%%%%%%%%%%%%%%%%%%%%%%%%%%%%%%%%%%%%%%%%%%%%%%%%%%%%%%%%%%%%%
%%%%%%%%%%%%%%%%%%%%%%%%%%%%%%%%%%%%%%%%%%%%%%%%%%%%%%%%%%%%%%%%%%%%%%%%%%%%%%%%

Since our regions were selected based on their angular sizes and not
their physical sizes, these results are unsurprising.  Because of our
angular size selection criterion, we would expect that the
nearer regions of the third quadrant are over-represented in our
sample, whereas the more distant regions of the first-quadrant outer
Galaxy are under-represented.  Because we lack distances to a
  high fraction of the sample (see Section~\ref{sec:dist}), we cannot test
  this hypothesis rigorously.

\begin{figure}
  \begin{centering}
  \includegraphics[width=3.25in]{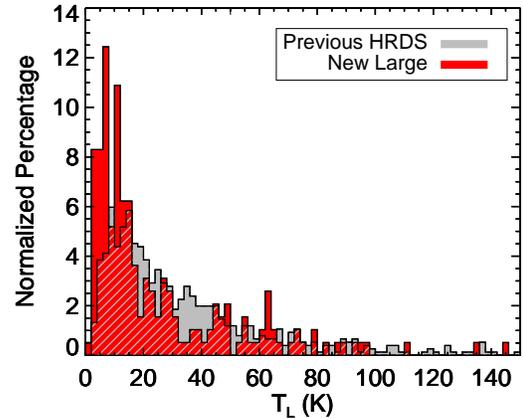}
  \caption{Normalized \hii\ region hydrogen RRL peak line intensity
    distributions of new large (red) and previous HRDS (gray)
    \hii\ regions.  Hatched areas show overlapping distributions. The
    previous HRDS measurements were conducted at X-band. Compared with
    the emission at C-band, optically thin X-band RRL emission that
    fills the telescope beam should have line intensities $\sim 50\%$
    as bright.  Instead, the large \hii\ region line intensities are
    substantially lower, which indicates that the sample contains many
    truly low surface brightness regions.\label{fig:peak}}
  \end{centering}
  \end{figure}

The distribution of hydrogen RRL peak line intensities for the new
large region sample is shifted toward lower values compared with the
previously-observed HRDS sample (Figure~\ref{fig:peak}).  Shown in
this figure (and subsequent calculations) are the intensities of all
hydrogen lines detected by the GBT, including all lines from multiple
RRL sources.  Previous HRDS surveys were conducted at X-band rather
than C-band. For optically thin \hii\ regions the observed antenna
temperature is roughly proportional to the observing frequency.  Based
on this effect alone, the C-band RRLs should be brighter than the
X-band RRLs by a factor of $\sim 2$.  Instead, the mean peak line
intensity for the new large regions is 25\,\mK\ with a standard
deviation of 26\,\mK, whereas for previous X-band HRDS surveys the
values are 45\,\mK\ and 111\,\mK, respectively.  This low average peak
intensity value shows that the new sample contains many truly low
surface brightness regions.

%%%%%%%%%%%%%%%%%%%%%%%%%%%%%%%%%%%%%%%%%%%%%%%%%%%%%%%%%%%%%%%%%%%%%%%%%%%%%%%%
%%%%%%%%%%%%%%%%%%%%%%%%%%%%%%%%%%%%%%%%%%%%%%%%%%%%%%%%%%%%%%%%%%%%%%%%%%%%%%%%

Of the 148 detections, 36 have spectra with multiple hydrogen RRL
components at different velocities.  In total, we detect emission
from 201 hydrogen RRL components (112 sources with one line, 19 with
two lines, and 17 with three lines).  As in \citet{anderson15b}, we
hypothesize that one of these components is from the discrete
\hii\ region that we targeted and the other(s) are from diffuse ionized
gas along the line of sight.
%To compute kinematic distances, we must determine the
%correct RRL velocity for these multiple-velocity \hii\ region spectra.
% copied from previous
In \citet{anderson15b} we derived a set of criteria that can be used
to determine which RRL component is from the discrete \hii\ region and
which component originates from diffuse gas.  Unfortunately, the lack
of continuum data here restricts our ability to use some of these
criteria.  The application of the remaining criteria was not
successful in determining which component arises from discrete
\hii\ regions.  Because of this, we are unable to determine which
component arises from the discrete \hii\ regions.

\begin{figure}
  \begin{centering}
  \includegraphics[width=3.25in]{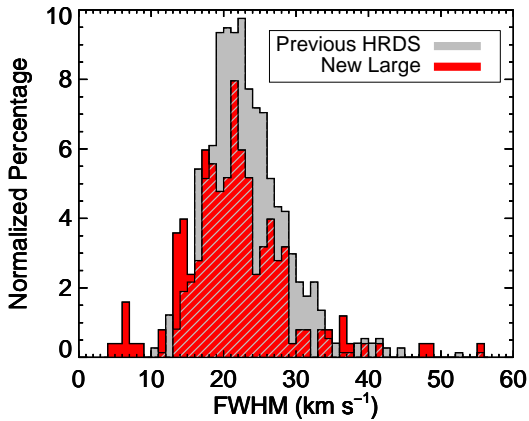}
  \caption{Normalized \hii\ region hydrogen RRL FWHM line width
    distributions of new large (red) and previous HRDS (gray)
    \hii\ regions. Although the shapes of the distributions are
    similar, there is a higher percentage of the new large regions
    that have hydrogen RRL line widths $<15\,\kms$, including seven
    new large regions with FWHM line widths
    $<10\,\kms$.\label{fig:fwhm}}
\end{centering}
\end{figure}

The RRL FWHM line widths of the current sample are lower on average
compared with that of previous HRDS surveys.  We show in
Figure~\ref{fig:fwhm} histograms of the current sample of large
regions (red) and those of previous HRDS surveys catalog (grey).  The
average FWHM for the new large regions is $21.6\pm 7.2\,\kms$, whereas
it is $23.3 \pm 5.3\,\kms$ for the sample from previous X-band HRDS
surveys.  Although these values are similar within the standard
deviations of the sample, a Kolmogorov-Smirnov (K-S) test shows that
the two FWHM samples are statistically distinct.  The difference is
entirely due to the large regions with FWHM values $<10\,\kms$ (see
below).  If these are removed, the K-S test shows that the samples are
not distinct.

%%%%%%%%%%%%%%%%%%%%%%%%%%%%%%%%%%%%%%%%%%%%%%%%%%%%%%%%%%%%%%%%%%%%%%%%%%%%%%%%
%%%%%%%%%%%%%%%%%%%%%%%%%%%%%%%%%%%%%%%%%%%%%%%%%%%%%%%%%%%%%%%%%%%%%%%%%%%%%%%%

\begin{figure}
  \begin{centering}
  \includegraphics[width=3.25in]{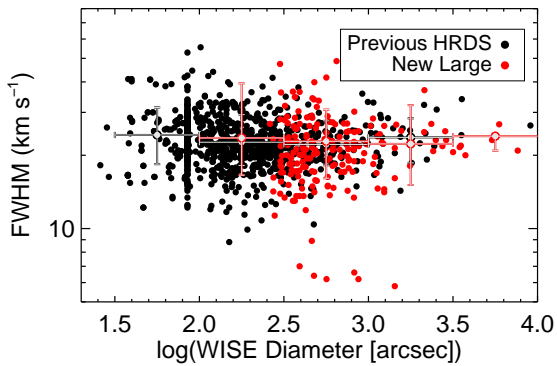}
  \caption{\hii\ region hydrogen RRL FWHM distributions of new large
    (red) and previous HRDS (black) \hii\ regions, as a function of
    angular diameter as defined in the {\it WISE} Catalog.  Open
    circles show averages in x-axis bins of 0.5. There is no
    relationship between the angular diameter and the RRL FWHM.  The
    many regions with diameters near $85\arcsec$ are an artifact due
    to how these sizes were defined in the {\it WISE}
    catalog; the true diameters for most such regions are slightly less than $85\arcsec$.. \label{fig:size_fwhm}}
\end{centering}
\end{figure}

Our sample is large in angular size compared with that of previous
HRDS surveys and a higher fraction of the nebulae have low RRL FWHM
line widths. There is, however, no correlation between the angular
size and the RRL FWHM.  We therefore do not believe that these two
quantities are causal. In Figure~\ref{fig:size_fwhm} we show the RRL
FWHM values as a function of the {\it WISE} Catalog-defined angular
diameter.  Note that many large regions were contained in previous
HRDS surveys, although they were not identified as such.

%%%%%%%%%%%%%%%%%%%%%%%%%%%%%%%%%%%%%%%%%%%%%%%%%%%%%%%%%%%%%%%%%%%%%%%%%%%%%%%%
%%%%%%%%%%%%%%%%%%%%%%%%%%%%%%%%%%%%%%%%%%%%%%%%%%%%%%%%%%%%%%%%%%%%%%%%%%%%%%%%

\begin{figure}
  \begin{centering}
  \includegraphics[width=3.25in]{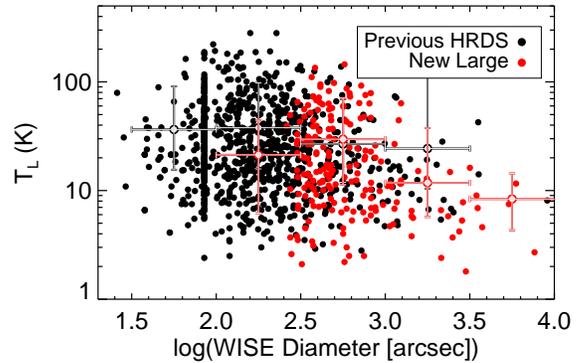}
  \caption{\hii\ region hydrogen RRL peak line height distributions of
    new large (red) and previous HRDS (black) \hii\ regions, as a
    function of angular diameter as defined in the {\it WISE} Catalog.
    Open circles show averages in x-axis bins of 0.5.  There is a weak
    trend between the angular diameter and the RRL peak line intensity
    such that nearly all the brightest regions with line intensities
    $\gtrsim 30$\,mK have angular diameters $<1000\arcsec$.  The many
    regions with diameters near $85\arcsec$ are an artifact due to how
    \hii\ region sizes were defined in the {\it WISE}
    Catalog; the true diameters for most such regions are slightly less than $85\arcsec$.. \label{fig:size_tl}}
\end{centering}
\end{figure}

There is a weak relationship between the peak RRL line intensity and
the {\it WISE} Catalog-defined angular diameter, as shown in
Figure~\ref{fig:size_tl}, such that the line intensity is lower on
average for the larger regions.  The relationship, although weak, is
expected.  \hii\ regions expand as they evolve.  Because the ionizing
photon flux from OB stars is roughly constant throughout their main
sequence lifetimes, their total ionized gas content is relatively
constant.  Therefore, their RRL intensities should decrease as their
physical sizes increase.

%%%%%%%%%%%%%%%%%%%%%%%%%%%%%%%%%%%%%%%%%%%%%%%%%%%%%%%%%%%%%%%%%%%%%%%%%%%%%%%%
%%%%%%%%%%%%%%%%%%%%%%%%%%%%%%%%%%%%%%%%%%%%%%%%%%%%%%%%%%%%%%%%%%%%%%%%%%%%%%%%

\subsection{Narrow Line Width Nebulae}
\begin{figure*}
  \begin{centering}
  \includegraphics[width=3in]{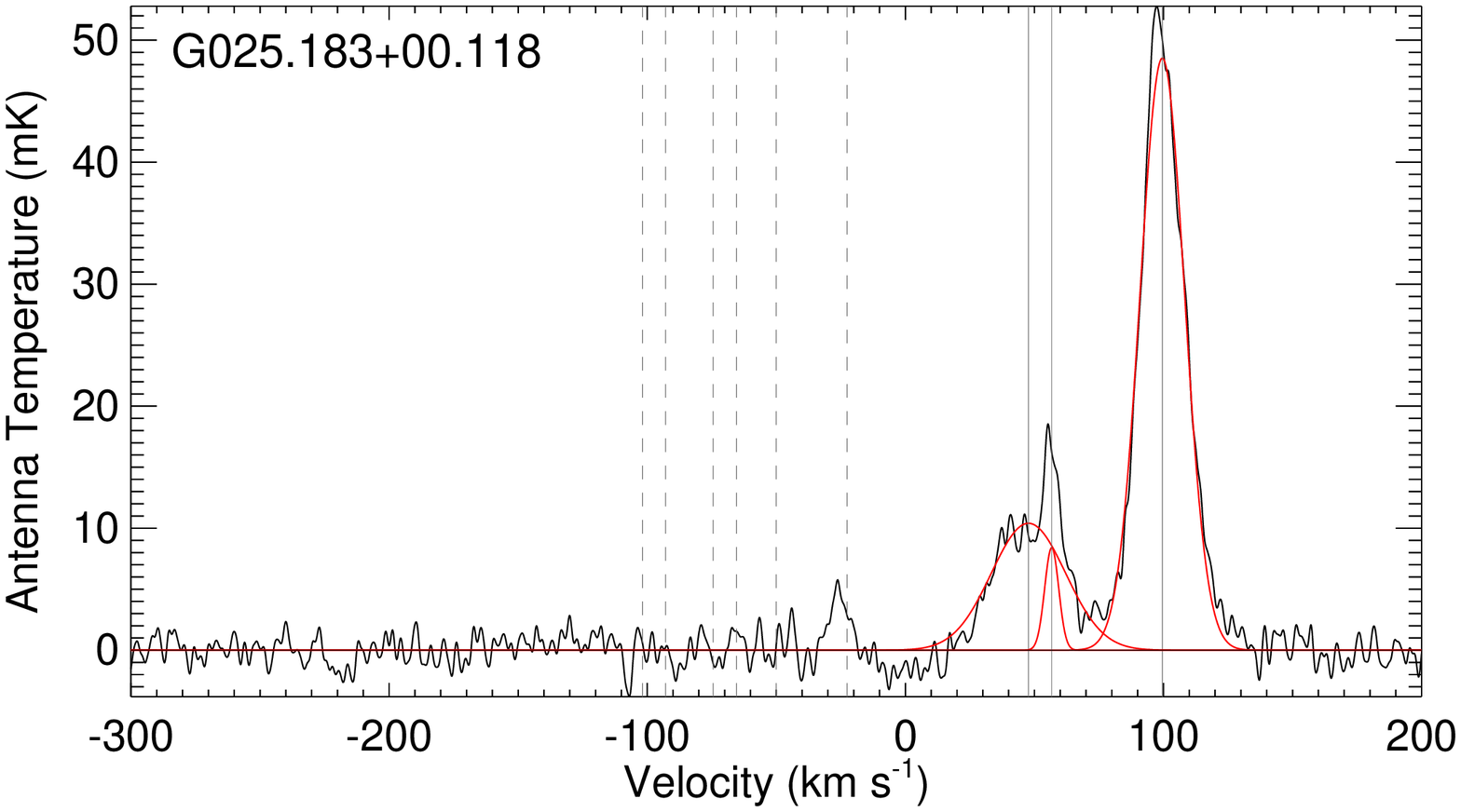}
  \includegraphics[width=3in]{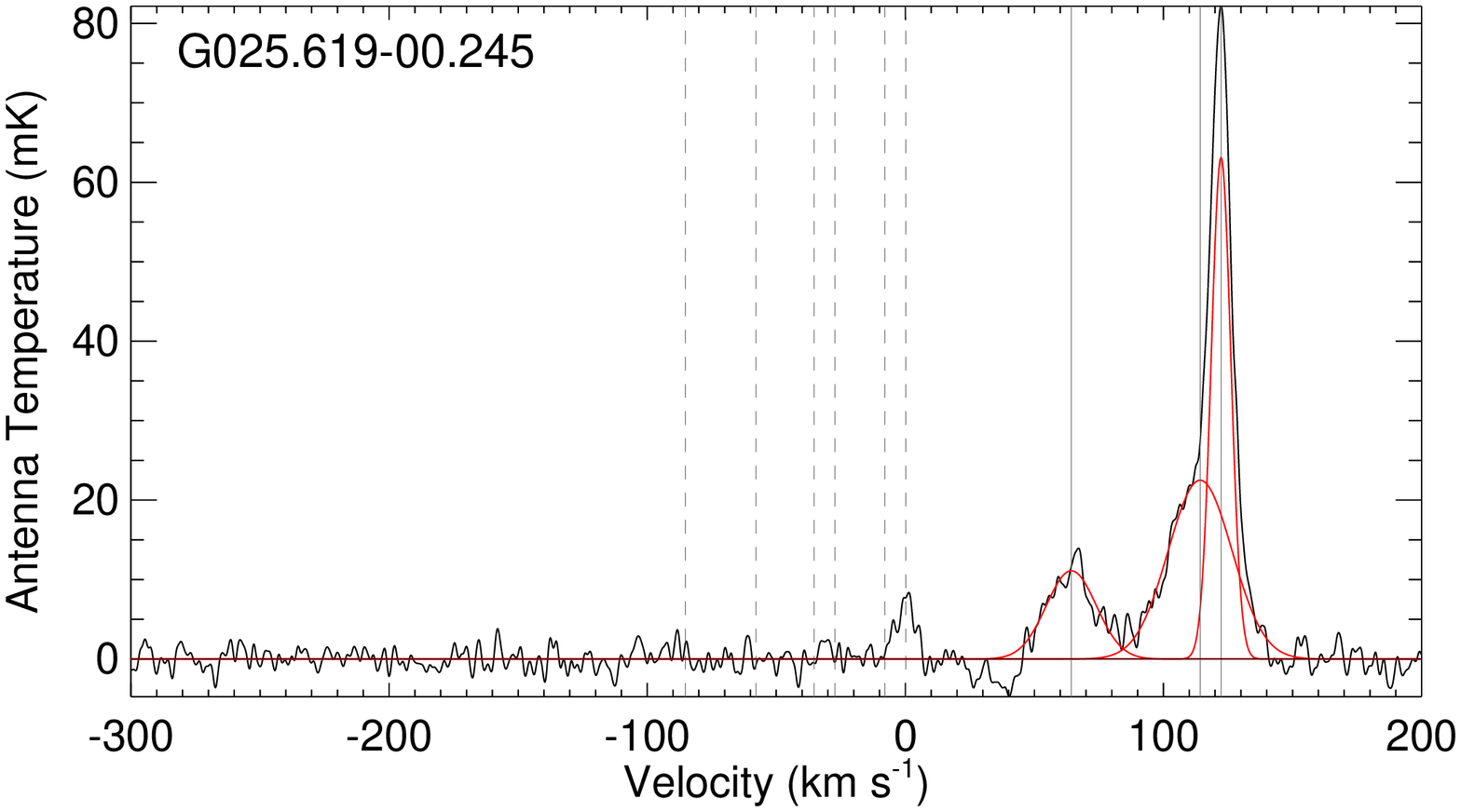}
  \includegraphics[width=3in]{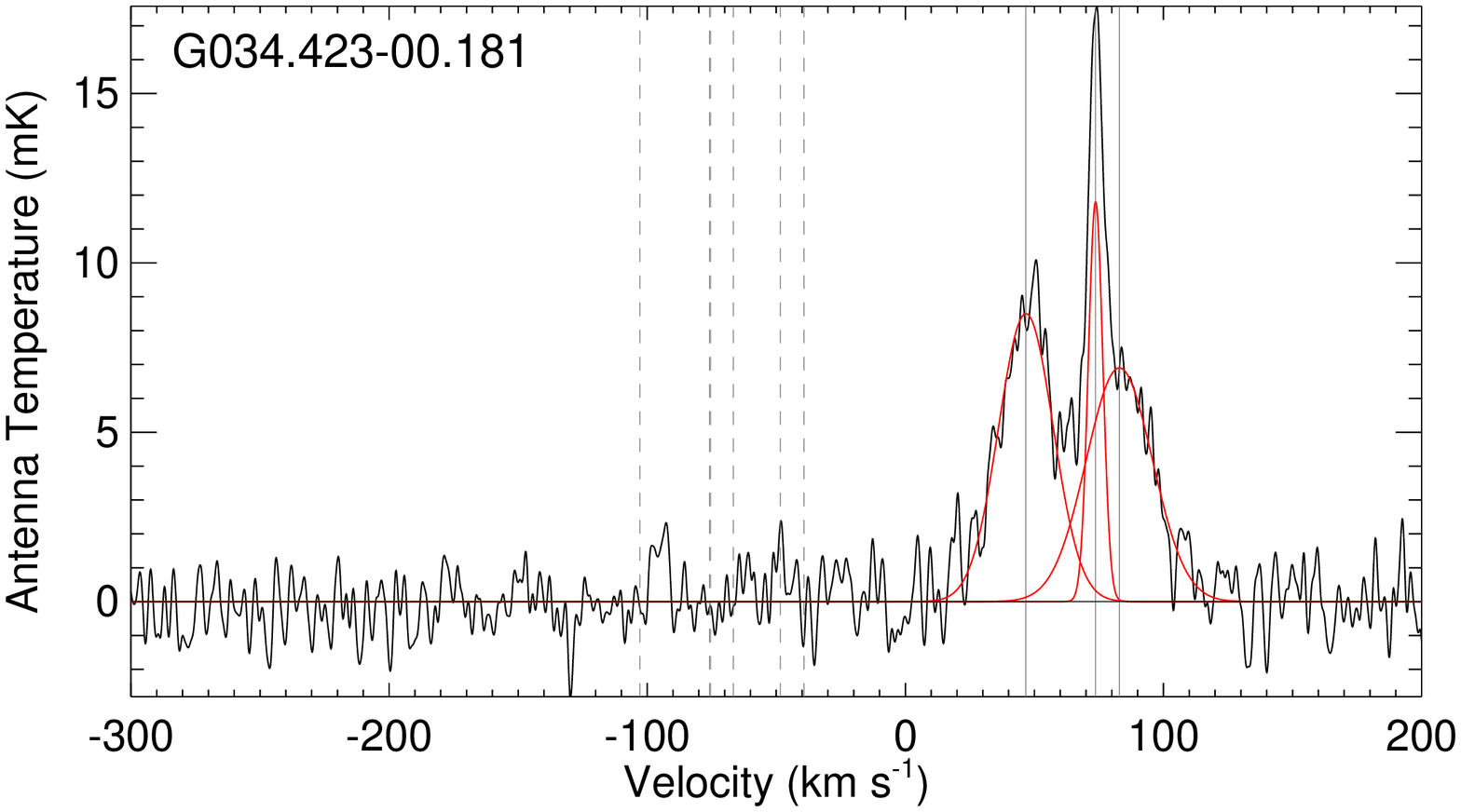}
  \includegraphics[width=3in]{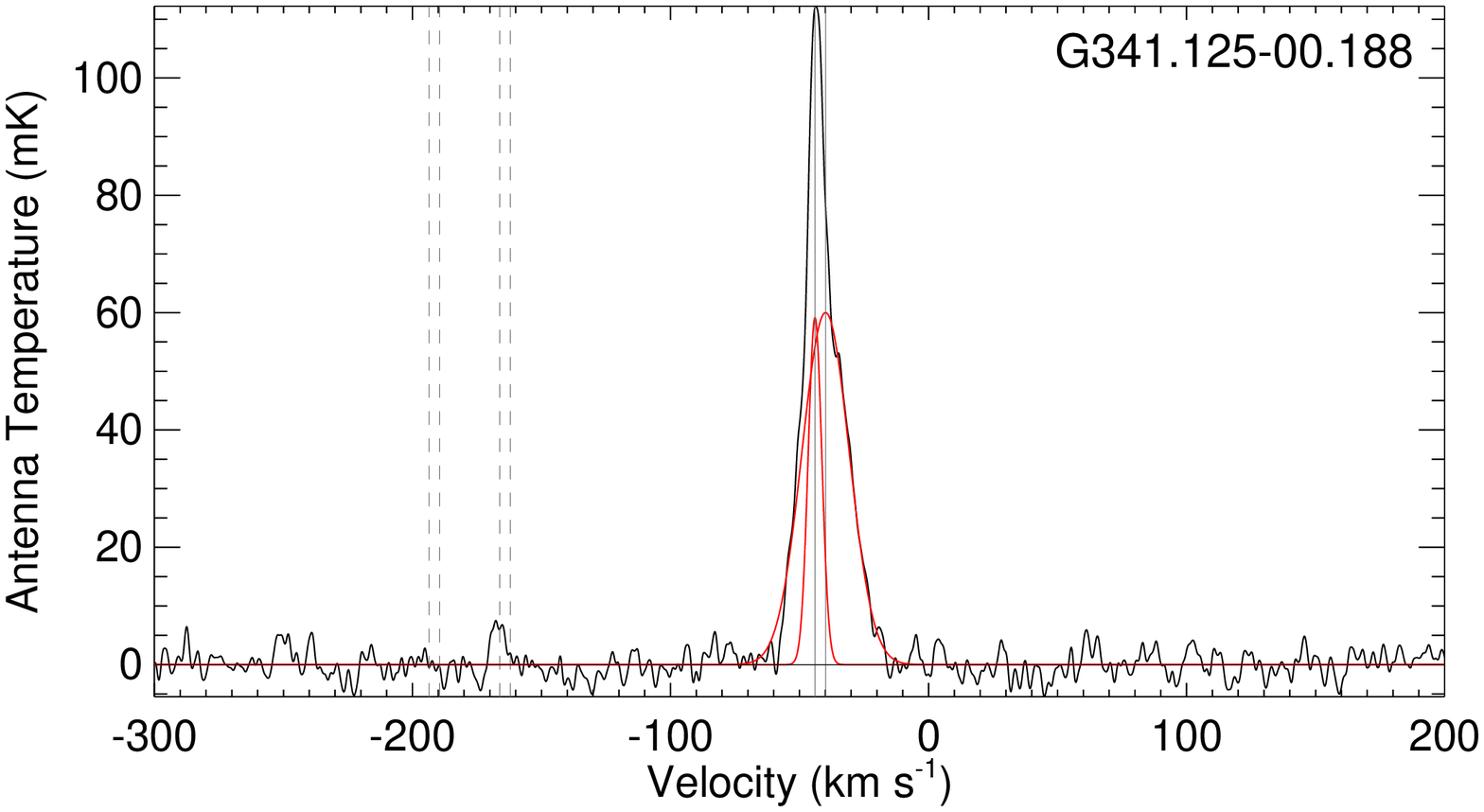}
  \includegraphics[width=3in]{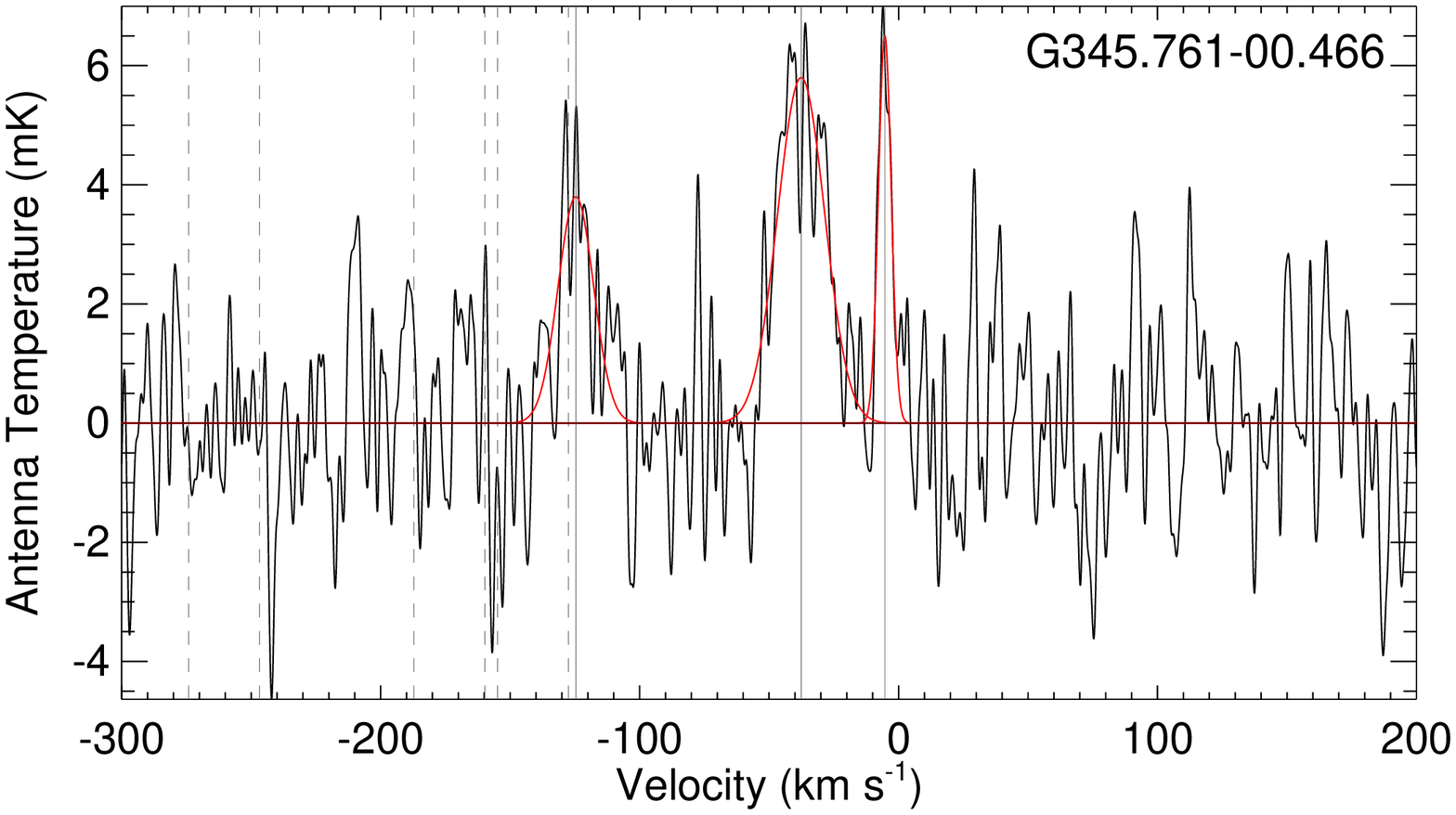}
  \includegraphics[width=3in]{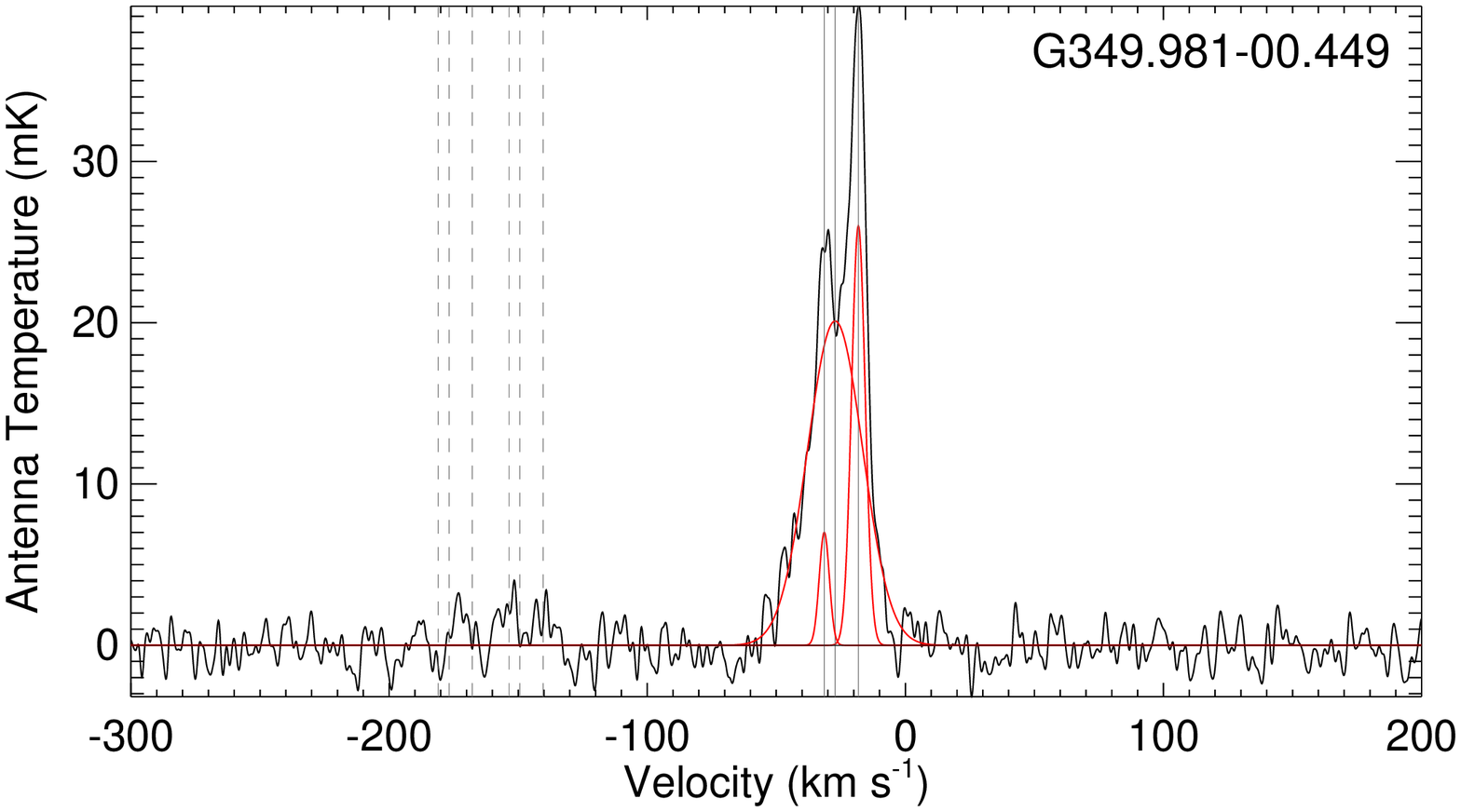}
  \includegraphics[width=3in]{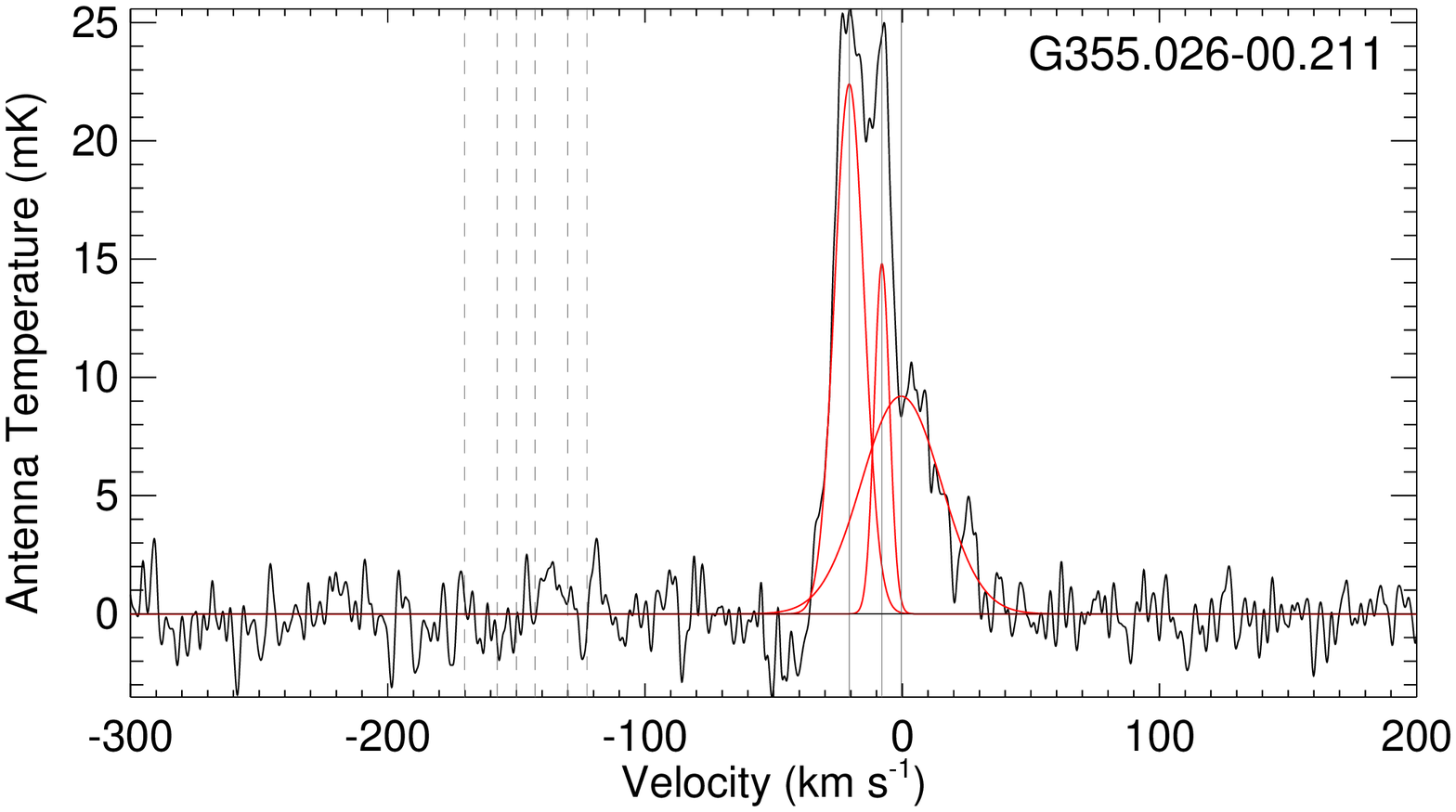}
  \caption{Spectra for nebulae with narrow hydrogen RRLs with line
    widths $<10\,\kms$.  The spectra are averages of lines
    H114$\alpha$ to H97$\alpha$, smoothed to 1.82\,\kms. Solid
    vertical lines show the fitted velocities of the hydrogen RRLs;
    dashed vertical lines show the expected velocities of the helium
    and carbon RRLs.
    \label{fig:narrow}}
  \end{centering}
\end{figure*}

\begin{figure}
  \begin{centering}
    \includegraphics[height=3.5in]{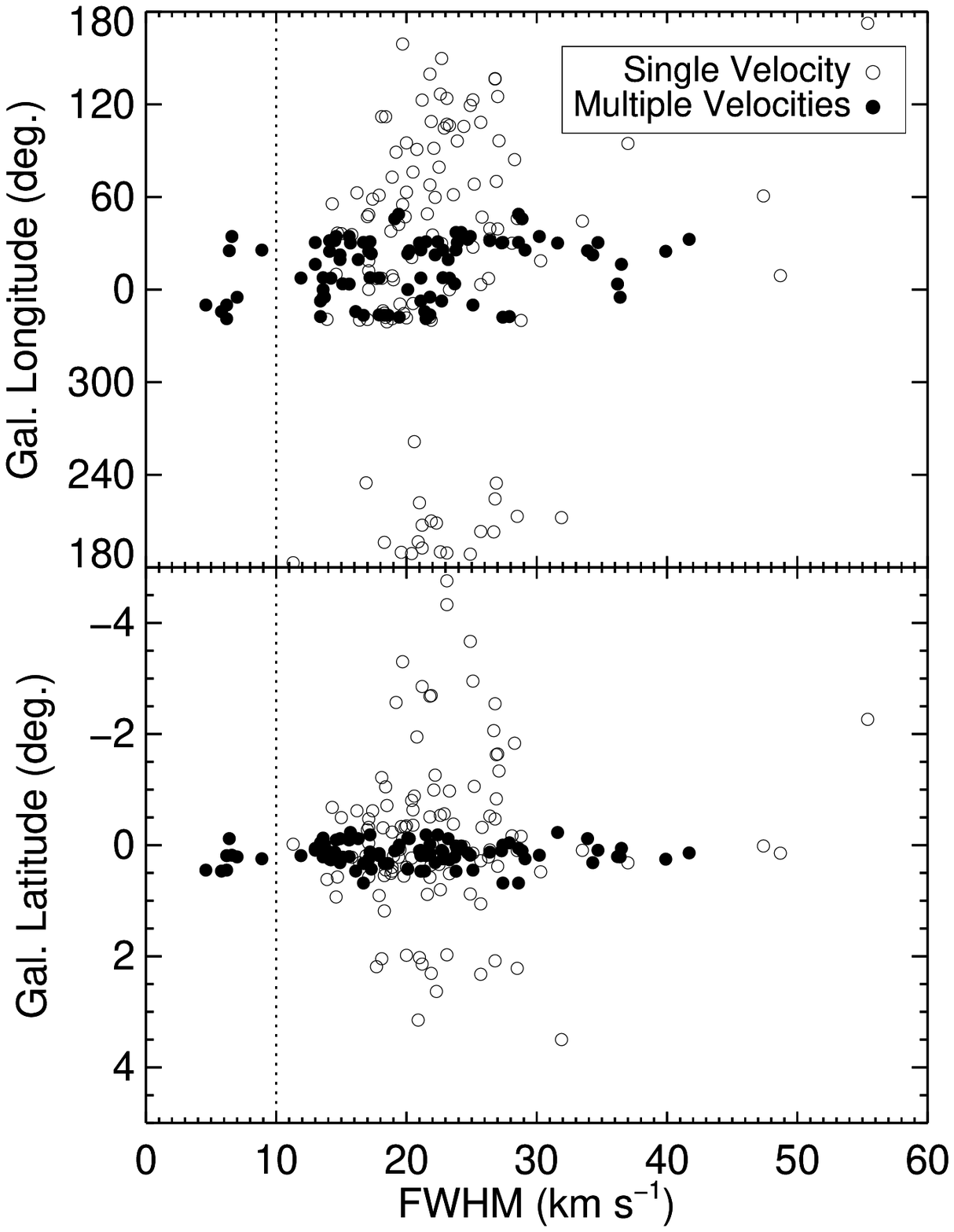}
  \caption{The distribution of Galactic longitudes (top) and Galactic
    latitudes (bottom) as a function of RRL FWHM values from this
    survey.  The narrow line width nebulae with FWHM~$<10\,\kms$
    (dotted line) are concentrated toward the inner Galaxy, near the
    mid-plane, as are the multiple-velocity
    nebulae.\label{fig:glgb_narrow}}
\end{centering}
\end{figure}

We detect the RRL emission from seven nebulae with FWHM line widths
$<10\,\kms$: G025.183+00.118, G025.619$-$00.245, G034.423$-$00.181,
G341.125$-$00.188, G345.761$-$00.466, G349.981$-$00.449 (two narrow
RRL components), and G355.026$-$00.211.  The spectra for all seven
sources, shown in Figure~\ref{fig:narrow}, have multiple RRL
components.  For all but G345.761$-$00.466, the narrow component is
blended with a broader component.  Additionally, six of the seven
nebulae have three velocity components, with only G341.125$-$00.188
showing just two.  In Figure~\ref{fig:glgb_narrow}, we show that all
narrow-line sources are located toward in the inner Galaxy at low
Galactic latitudes.  The infrared morphologies of these seven sources
are unremarkable; they look similar to other regions discovered in the
HRDS.

%%%%%%%%%%%%%%%%%%%%%%%%%%%%%%%%%%%%%%%%%%%%%%%%%%%%%%%%%%%%%%%%%%%%%%%%%%%%%%%%
%\begin{figure}
%  \includegraphics[width=3in]{../../hrds_wise/figures/spectra/FQ236.eps}
%  \includegraphics[width=3in]{../../hrds_wise/figures/spectra/LA574.eps}
%  \includegraphics[width=3in]{../../hrds_wise/figures/spectra/FQ016.eps}
%  \includegraphics[width=3in]{../../hrds_wise/figures/spectra/GS008.eps}
%  \includegraphics[width=3in]{../../hrds_survey/figures/spectra/LA563.eps}
%\end{figure}
%%%%%%%%%%%%%%%%%%%%%%%%%%%%%%%%%%%%%%%%%%%%%%%%%%%%%%%%%%%%%%%%%%%%%%%%%%%%%%%%

%%%%%%%%%%%%%%%%%%%%%%%%%%%%%%%%%%%%%%%%%%%%%%%%%%%%%%%%%%%%%%%%%%%%%%%%%%%%%%%%
%%%%%%%%%%%%%%%%%%%%%%%%%%%%%%%%%%%%%%%%%%%%%%%%%%%%%%%%%%%%%%%%%%%%%%%%%%%%%%%%

For each narrow-line source, we examine the individual
single-polarization spectra H114$\alpha$ through H97$\alpha$ to verify
that the narrow lines in the combined spectra are not due to radio
frequency interference, instrumental artifacts, or some other
transition within the bandpass.  We fit Gaussian models to the
individual spectra if the narrow lines are bright enough.  The
r.m.s. noise in individual spectra at a single polarization is $\sim
10$\,\mK, so the lines must be $\gtrsim 25$\,\mK to be detectable.
This limits the analysis to G025.619$-$00.245, 341.125$-$00.188,
G349.981$-$00.449, and G355.026$-$00.211.  For the narrow velocity
components of these sources, the average derived FWHM values from the
individual spectra differ by at most 20\% from those from the combined
spectra.  This exercise provides confirmation that the narrow lines of
these four sources are real, and furthermore that the narrow lines are
seen in spectra taken with beam sizes from $110\arcsec$ to
$170\arcsec$.
%% As an additional chaeck against such issues, one
%% could also inspect the spectra for He lines offset by $-122\,\kms$
%% from the narrow H lines at $\sim 10\%$ of the H line intensities.
%% Unfortunately, the fact that nearly all narrow lines are blended with
%% broader lines and that many of the narrow lines are weak makes such an
%% analysis difficult.

%% Such narrow lines can be explained by 1)~a combination of low electron
%% temperature plasma and/or low turbulent broadening, 2)~if the detected
%% lines are actually due to carbon instead of hydrogen, or 3)~if the
%% hydrogen lines are masing.
If the line broadening is caused only by thermal motions, line widths
between 5 and 10\,\kms\ corresponds to electron temperatures between
500 and 2200\,\K.  If the thermal and turbulent broadening contribute
equally to the line widths, the thermal line widths are then between
3.5 and 7.1\,\kms, and the electron temperatures are between 300\,\K
and 1100\,\K.  We note that the lowest electron temperatures derived
using RRL measurements of Galactic \hii\ regions that have simple RRL
profiles fit with single Gaussian models are near 6000\,\K
\citep{balser11}, as are those derived using optical emission lines
\citep{deharveng00}.
%([OII] λλ3726 and 3729, [O III] λλ4363 and
  %  5007, He Iλ5876, Hα and Hβ)
% In Google: (7.1 km/s)^2 * 1amu / (8 * k * ln(2)) in K
%
% Equal parts gives 7.1km/s and 3.5km/s
%% If the ambient ISM pressure in the mid-plane is $p/k \simeq
%% 22,000\,\percc\,\K$ \citep{cox05} and the regions are in pressure
%% equilibrium with the ISM, these regions must have densities ranging
%% from 15 to 4\,\percc for the case of no turbulence, and 30 to
%% 8\,\percc when half the line width is due to turbulence.  These values
%% are not extraordinary, and so by themselves do not exclude such low
%% plasma temperatures.
%% % 3000 / 1500
%% {\ldacom Lower Te for lower ne?  Get Dana's reference.}

There are many possible explanations for such narrow RRLs.  Similarly
narrow RRL profiles have been found at multiple positions in the W33
complex \citep{bieging78}, and have been argued to exist as a diffuse
component of the Galactic disk \citep{lockman80}.  The authors
speculate that the narrow lines may be caused by interactions between
ionized and molecular gas. The narrow lines may also arise from
partially ionized hydrogen within the \hii\ region PDRs, as has been
discussed for W3 \citep{adler96}.  In the case of W3, however, the
line profiles for the partially ionized and fully ionized components
are at the same velocity, whereas here ours are all offset.  If these
lines do arise from partially-ionized zones, the PDRs would have to be
moving relative to the ionized gas.  The offset between hydrogen
(which traces the ionized gas) and carbon (which traces PDRs) RRL
velocities is $\sim 5\,\kms$ for Galactic \hii\ regions
\citep{wenger13}.  If the narrow lines are from partially ionized
zones, the relative motion of the ionized gas and the PDRs would be
exceptional for Galactic \hii\ regions.  Runaway OB stars could also
create large offsets between partially and fully ionized components,
although there is no indication from the source morphologies that
these regions are powered by runaway OB stars.  \citet{onello95}
observed very narrow RRLs ($\sim 3\,\kms$) toward the source G70.7+1.2
and concluded that the narrow lines may be caused by cold gas near the
outer boundary of a bow shock. This object is unusual in many respects
and this therefore cannot be a good explanation for the narrow lines
discovered here.

\begin{figure*}
  \begin{centering}
    \includegraphics[height=2in]{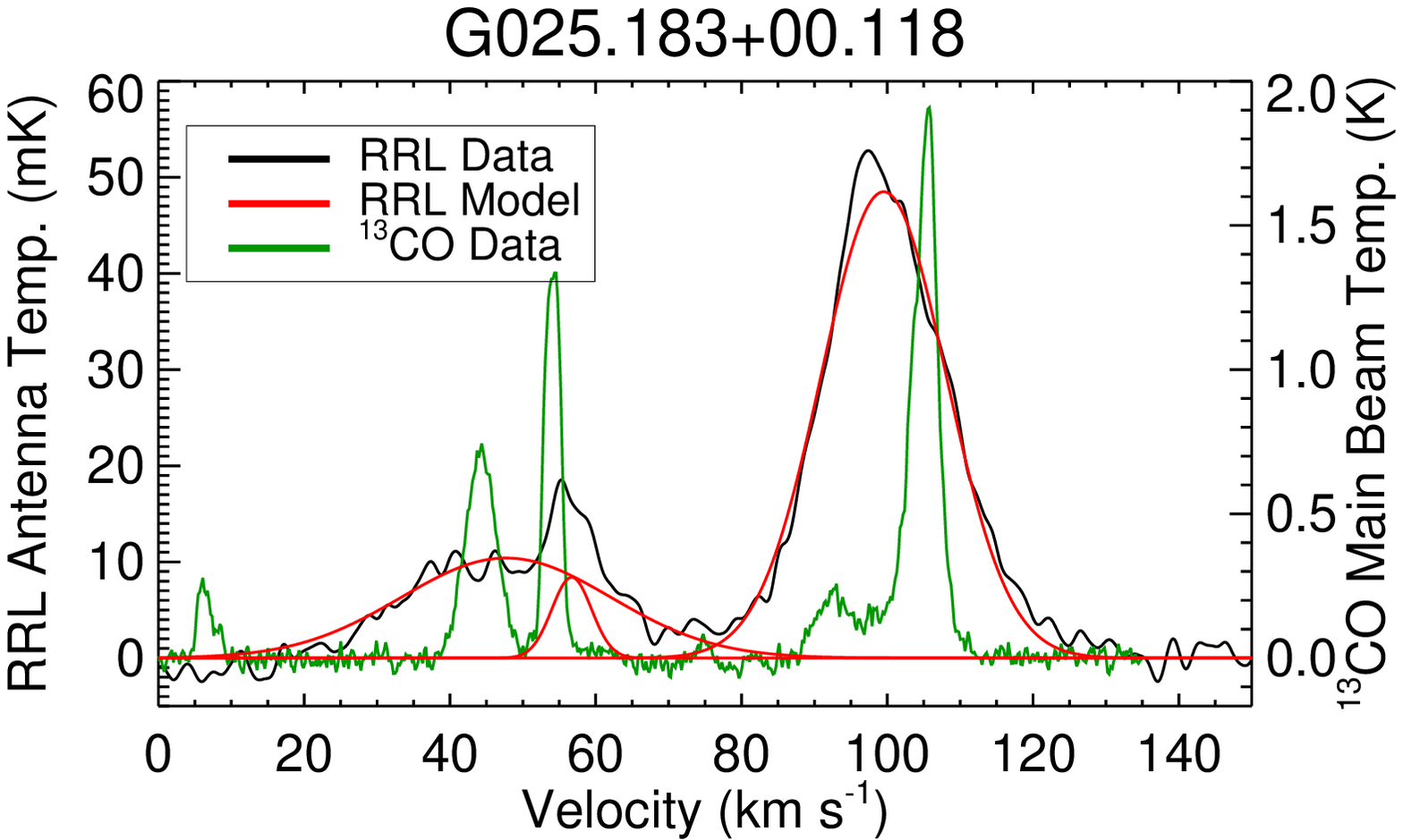}
    \includegraphics[height=2in]{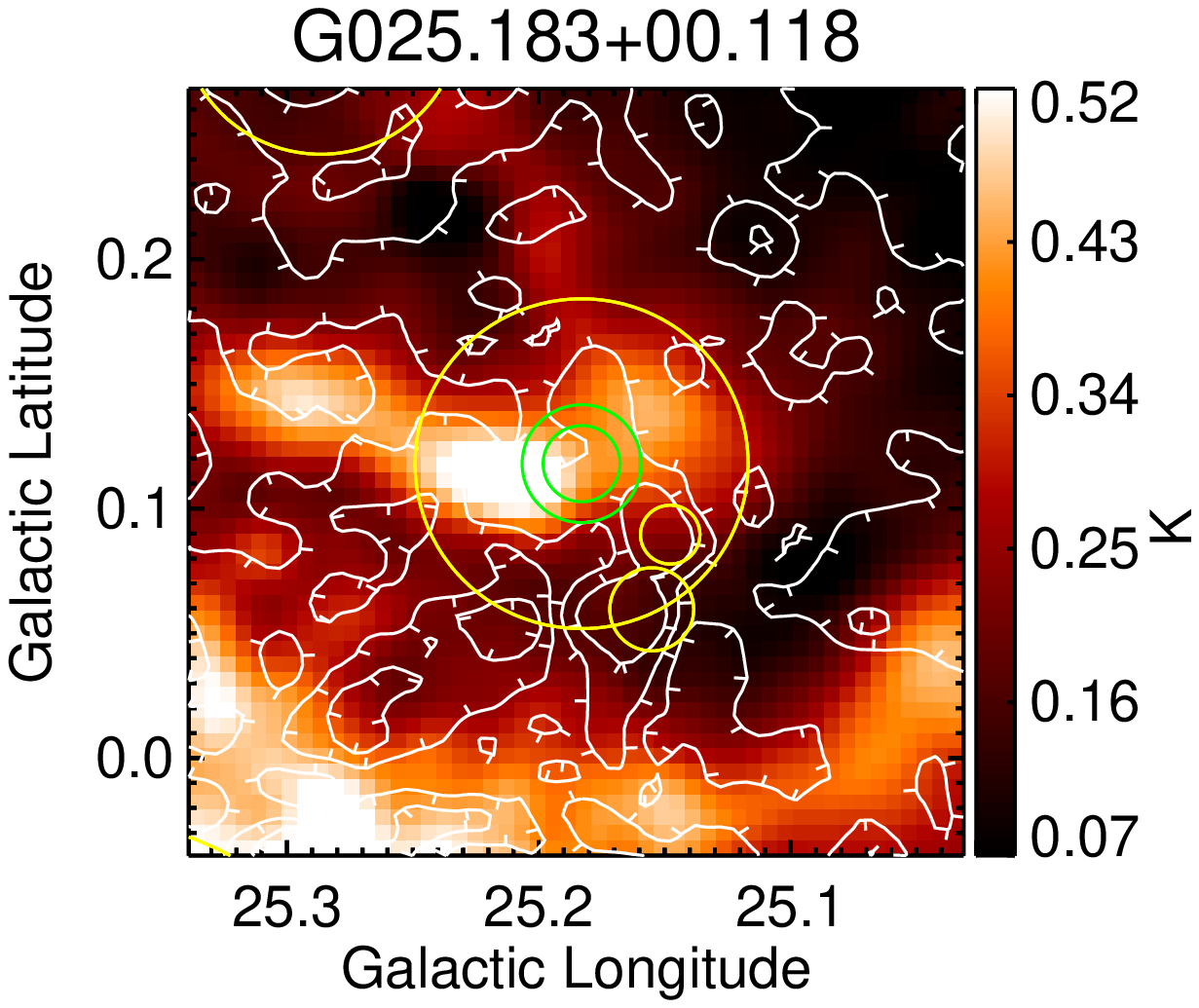}
    \includegraphics[height=2in]{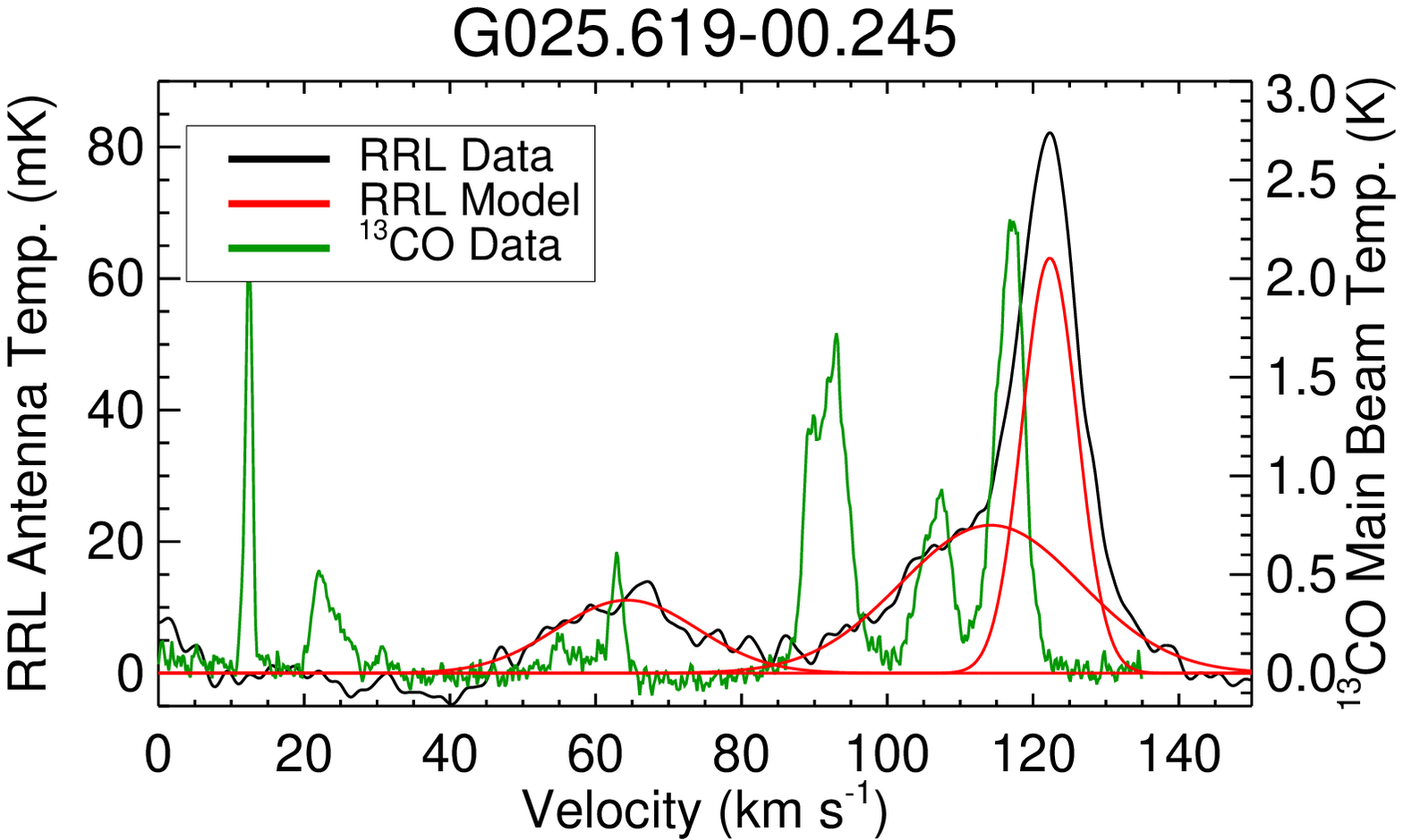}
    \includegraphics[height=2in]{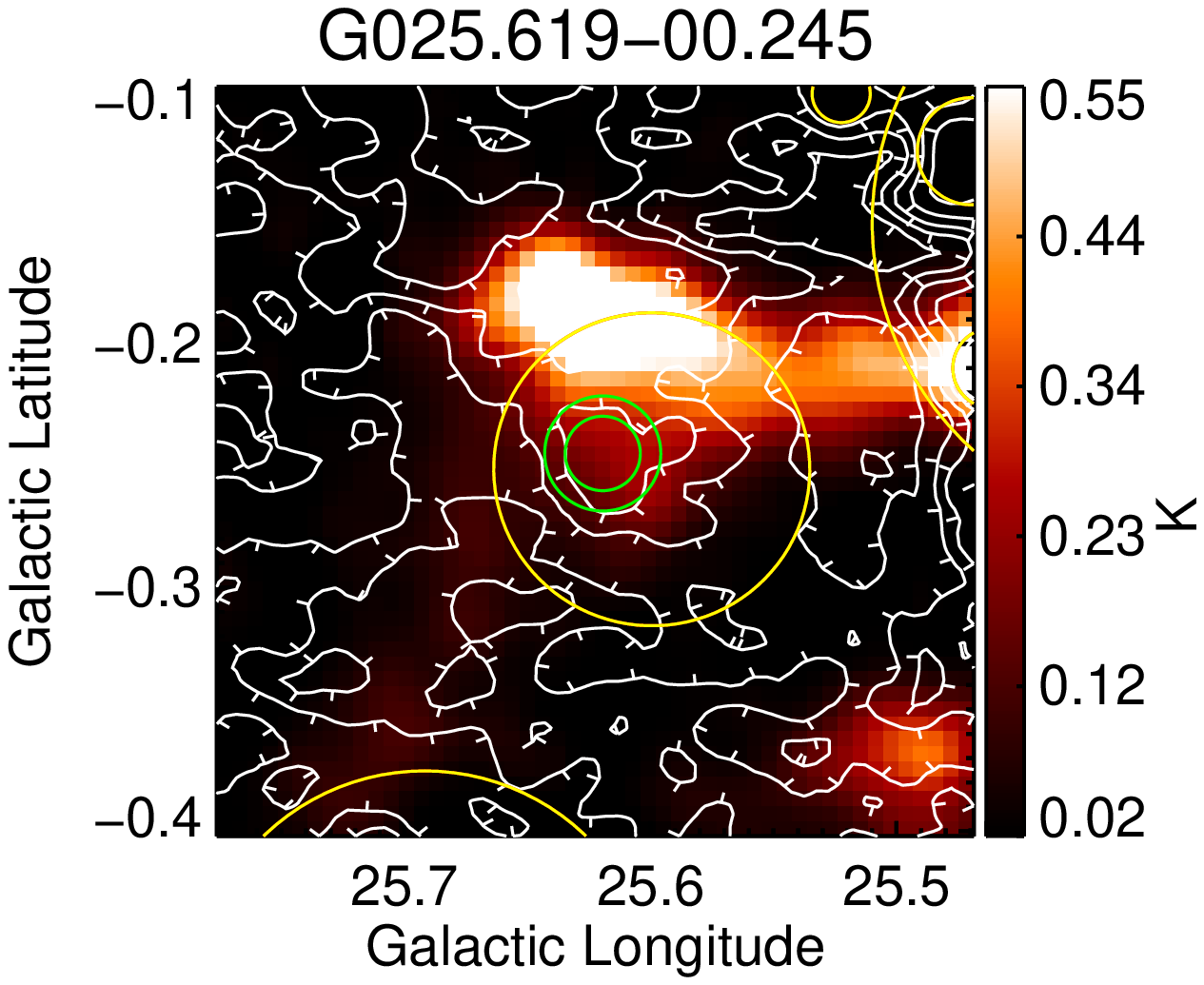}
    \includegraphics[height=2in]{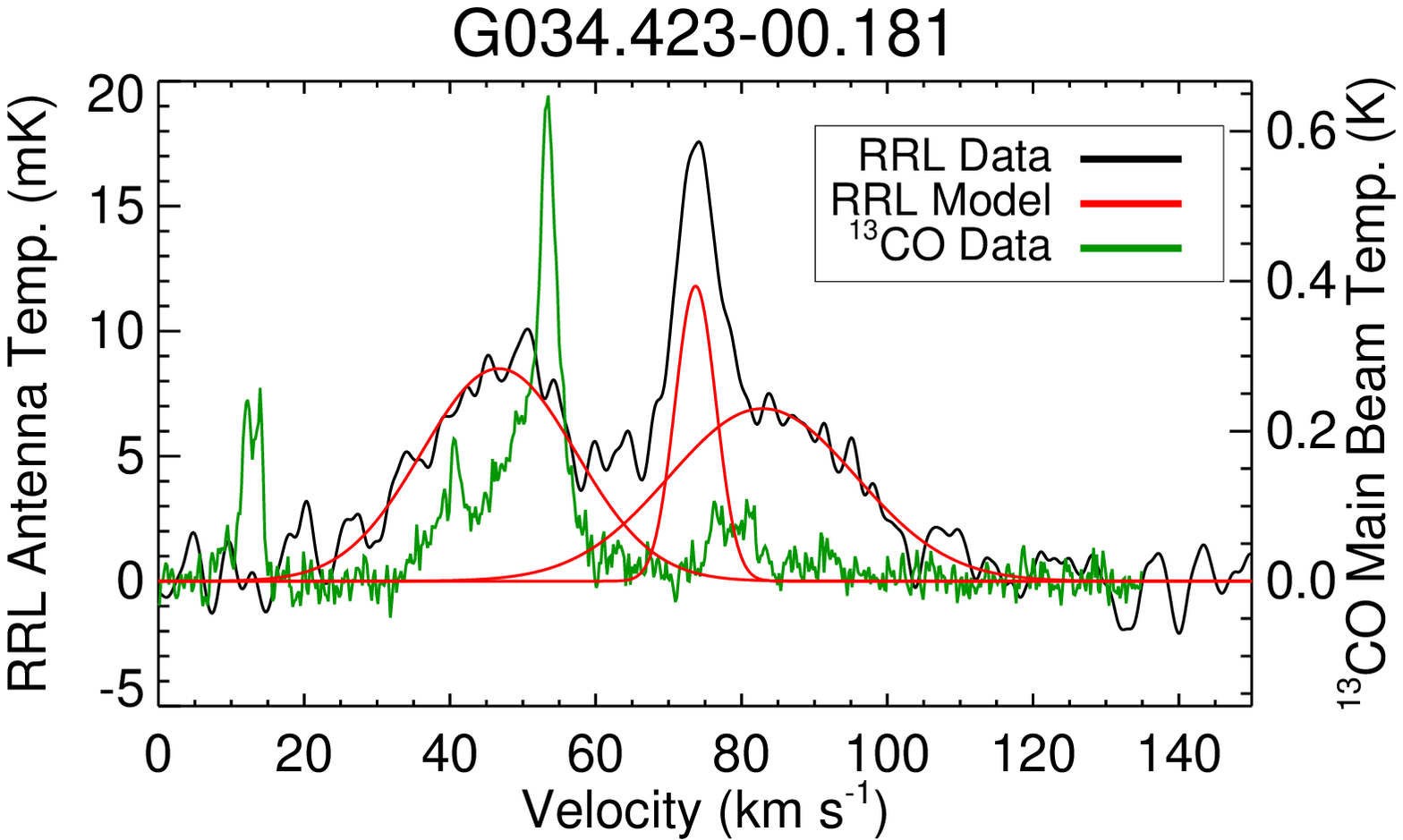}
    \includegraphics[height=2in]{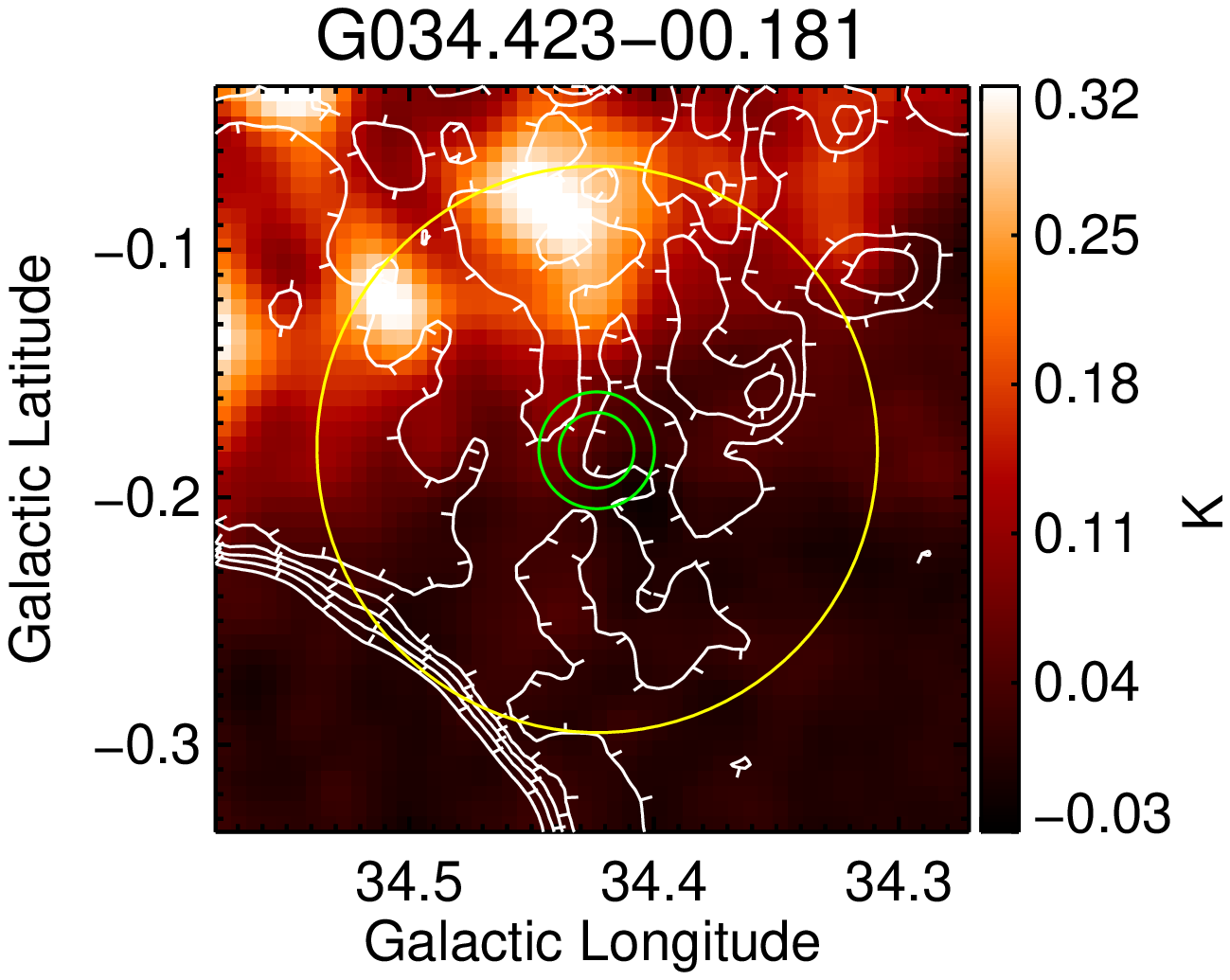}
  \caption{Galactic Ring Survey \cor\ spectra (left column) and
    position-position maps (right column) for three narrow-line
    \hii\ regions.  The position-position maps are centered on the
    observed location.  The CO emission is integrated over the narrow
    RRL line width.  The green circles show the GBT beam at the lowest
    and highest observed frequencies.  The yellow circle at the center
    shows the infrared extent of the targeted region; other yellow
    circles show other \hii\ regions in the field.  The white contours
    show VGPS 21\,cm continuum contours, with values of 18, 20, 22,
    25, and 30\,\K; ticks on the contour lines point
    downhill. Although there is \cor\ emission at all narrow RRL
    velocities, this emission is too complicated to definitively say
    that the narrow RRL velocities are related to emission from local
    molecular gas.\label{fig:narrow_grs}}
\end{centering}
\end{figure*}

We find that although there is molecular emission at the narrow-RRL
velocities, there is no strong indication that the
$^{13}$CO and narrow line components are related. 
%We favor the explanation that the narrow lines are somehow caused by
%interactions with molecular gas, perhaps as part of the \hii\ region
%PDRs, and therefore examine molecular line emission near the regions.
  %In addition to \hii\ region PDRs can emit strongly in molecular
%lines \citep{deharveng03}.
We examine $^{13}$CO data from the
Galactic Ring Survey \citep[GRS;][]{jackson06} for the three regions
within the GRS survey area: G025.183+00.118, G025.619$-$00.245, and
G034.423$-$00.181.  We show the \cor\ spectra and position-position
images of $^{13}$CO intensity in Figure~\ref{fig:narrow_grs}.  We
created the position-position images by averaging over the narrow RRL
velocity range.   The molecular
emission looks similar to that of other angularly-large \hii\ regions
analyzed previously in \citet{anderson09b}.  Furthermore, the observed
locations are far away from the PDRs enclosed by the red circles.
These data are too confusing to provide evidence on the subject of
whether the narrow-lines are coming from interactions with molecular
clouds or from partially ionized zones within the \hii\ region PDRs.

%%%%%%%%%%%%%%%%%%%%%%%%%%%%%%%%%%%%%%%%%%%%%%%%%%%%%%%%%%%%%%%%%%%%%%%%%%%%%%%%

Although we do not have a suitable explanation for these narrow-line
regions, one clue may be that they all have multiple hydrogen RRL
components and are located in the inner Galaxy.  This may be
coincidence, or may indicate that the narrow RRL components may be
caused in some way by the WIM.  The WIM also exists in the inner
Galaxy and is the source of many additional RRL components detected in
observations of \hii\ regions \citep{anderson15b}.
%The WIM is a
%low-density plasma that has its strongest emission in the inner
%Galaxy.  In \citet{anderson15b} we showed that many of the secondary
%and tertiary RRL line components detected in the HRDS are due to
%diffuse ionized gas, presumably from the WIM.
There were four narrow-line sources with FWHM~$<10\,\kms$ in the HRDS
survey of \citet{anderson15c}, and all four have multiple RRL line
components and exist in the inner Galaxy.  There was only one such
narrow RRL source in the original HRDS \citep{anderson11}, and it too
was a multiple RRL source in the inner Galaxy.
%Of the HRDS sources detected to date, roughly 25\% have had
%two line components, and 24\% of the present survey.  There is a
%$0.25^{12} = 6\times 10^{-6}\%$ chance of this occuring randomly.

%% It is important to note that these narrow lines are not seen in
%% detections of the WIM in directions devoid of discrete \hii\ regions.
%% For example, in \citet{luisi16}, we measured the diffuse RRL emission
%% escaping from the \hii\ region NGC7538, using GBT X-band observations.
%% The line width of the diffuse gas outside of NGC7538's PDR was 27.6
%% with a standard deviation of 3.8\,\kms, whereas it was 30.9 with a
%% standard deviation of 2.4\,\kms\ inside the PDR.  We propose that they
%% must somehow be caused b the interaction between the WIM and discrete
%% \hii\ regions.

%None of these lines were seen in the
%individual spectra, which indicates that they are hydrogen RRLs.

\subsection{Distances\label{sec:dist}}
We derive kinematic distances for 45 of the detected \hii\ regions, 35
of which lie in the outer Galaxy ($\rgal > 8.5\,\kpc$).  The results
of our kinematic distance analysis, in addition to the numbers of
regions known previously, are summarized in Table~\ref{tab:distances},
which lists the source name, the LSR velocity, the near, far, and
tangent point distances, the Galactocentric radius, the tangent point
velocity, the kinematic distance ambiguity resolution (``KDAR;'' ``T''
= tangent point distance, ``O'' = outer Galaxy), the Heliocentric
distance, the Heliocentric distance uncertainty computed from the
\citet{anderson14} model, and the vertical distance, $z$, from the
Galactic mid-plane.  The \citet{anderson14} distance uncertainty
  model includes contributions from streaming motions, the uncertainty
  in the ``true'' rotation curve model, and uncertainty in the Solar
  rotation parameters.

%%%%%%%%%%%%%%%%%%%%%%%%%%%%%%%%%%%%%%%%%%%%%%%%%%%%%%%%%%%%%%%%%%%%%%%%%%%%%%%%
\begin{deluxetable*}{lrcccccccrrr}
\setlength{\tabcolsep}{6pt}
\tabletypesize{\footnotesize}
\tablecaption{Kinematic Distances}
\tablewidth{0pt}
\tablehead{
\colhead{Name} &
\colhead{$V_{LSR}$} &
\colhead{$D_N$} &
\colhead{$D_F$} &
\colhead{$D_T$} &
\colhead{\rgal} &
\colhead{$V_{T}$} &
\colhead{KDAR$^{\rm a}$} &
\colhead{$D_\sun$} &
\colhead{$\sigma D_\sun$} &
\colhead{$z$}
\\
\colhead{} &
\colhead{(\kms)} &
\colhead{(kpc)} &
\colhead{(kpc)} &
\colhead{(kpc)} &
\colhead{(kpc)} &
\colhead{(\kms)} &
\colhead{} &
\colhead{(kpc)} &
\colhead{(kpc)} &
\colhead{(pc)}
}
\startdata
G029.828$-$00.167 & $101.4$ & 6.3 & \phn8.5 & 7.4 & \phn4.4 & 107.9 & T & \phn7.4 & 1.1 & \phn$-21$\\
G032.687$-$00.086 & $103.3$ & 7.2 & \phn7.2 & 7.2 & \phn4.6 & \phn99.2 & T & \phn7.2 & 0.0 & \phn$-10$\\
G039.515+00.523 & \phn$-30.8$ & \nodata & 16.3 & 6.6 & 11.2 & \phn79.4 & O & 16.3 & 1.2 & 149\\
G044.365$-$00.092 & \phn$56.7$ & 4.2 & \phn7.9 & 6.1 & \phn6.2 & \phn66.3 & T & \phn6.1 & 1.9 & \phn\phn$-9$\\
G046.033$-$00.096 & \phn$63.5$ & 5.9 & \phn5.9 & 5.9 & \phn6.1 & \phn62.1 & T & \phn5.9 & 0.0 & \phn\phn$-9$\\
G046.964+00.320 & \phn$-41.7$ & \nodata & 15.8 & 5.8 & 11.8 & \phn59.7 & O & 15.8 & 1.2 & \phn88\\
G047.181+00.326 & \phn$52.0$ & 4.1 & \phn7.4 & 5.8 & \phn6.5 & \phn59.2 & T & \phn5.8 & 1.7 & \phn32\\
G047.280+00.276 & \phn$59.9$ & 5.8 & \phn5.8 & 5.8 & \phn6.2 & \phn58.9 & T & \phn5.8 & 0.0 & \phn27\\
G049.049$-$00.886 & \phn$54.3$ & 5.2 & \phn5.9 & 5.6 & \phn6.4 & \phn54.6 & T & \phn5.6 & 0.4 & \phn$-86$\\
G055.111+00.111 & \phn$31.9$ & 2.8 & \phn7.0 & 4.9 & \phn7.3 & \phn41.0 & T & \phn4.9 & 2.1 & \phn\phn9\\
G055.585+00.680 & \phn\phn$-0.2$ & \nodata & \phn9.8 & 4.8 & \phn8.6 & \phn40.0 & O & \phn9.8 & 0.9 & 116\\
G058.607+00.619 & \phn$29.8$ & 3.0 & \phn5.9 & 4.4 & \phn7.4 & \phn34.0 & T & \phn4.4 & 1.5 & \phn47\\
G059.747+01.260 & \phn$-50.3$ & \nodata & 13.6 & 4.3 & 11.9 & \phn31.9 & O & 13.6 & 1.2 & 299\\
G060.654$-$00.015 & \phn\phn$-5.5$ & \nodata & \phn9.1 & 4.2 & \phn8.9 & \phn30.2 & O & \phn9.1 & 1.0 & \phn\phn$-2$\\
G061.467+00.380 & \phn$23.1$ & 2.3 & \phn5.8 & 4.1 & \phn7.7 & \phn28.8 & T & \phn4.1 & 1.8 & \phn26\\
G067.814+00.509 & \phn\phn$-3.1$ & \nodata & \phn7.1 & 3.2 & \phn8.8 & \phn18.8 & O & \phn7.1 & 1.2 & \phn62\\
G070.099+01.629 & \phn$-32.6$ & \nodata & \phn9.4 & 2.9 & 10.3 & \phn15.8 & O & \phn9.4 & 1.3 & 267\\
G072.845+00.235 & \phn$18.2$ & 2.5 & \phn2.5 & 2.5 & \phn8.1 & \phn12.6 & T & \phn2.5 & 0.1 & \phn10\\
G090.842+01.948 & \phn$-83.1$ & \nodata & 11.2 & \nodata & 14.1 & \nodata & O & 11.2 & 1.4 & 379\\
G091.546+00.993 & \phn$-54.0$ & \nodata & \phn7.6 & \nodata & 11.6 & \nodata & O & \phn7.6 & 1.4 & 131\\
G094.665$-$00.318 & \phn$-38.9$ & \nodata & \phn5.6 & \nodata & 10.6 & \nodata & O & \phn5.6 & 1.4 & \phn$-31$\\
G095.035$-$01.982 & \phn$-45.7$ & \nodata & \phn6.3 & \nodata & 11.0 & \nodata & O & \phn6.3 & 1.4 & $-217$\\
G096.310$-$00.079 & \phn$-47.9$ & \nodata & \phn6.3 & \nodata & 11.2 & \nodata & O & \phn6.3 & 1.4 & \phn\phn$-8$\\
G096.415+01.335 & \phn$-67.5$ & \nodata & \phn8.5 & \nodata & 12.7 & \nodata & O & \phn8.5 & 1.2 & 197\\
G104.662+00.564 & \phn$-71.1$ & \nodata & \phn8.1 & \nodata & 13.2 & \nodata & O & \phn8.1 & 1.3 & \phn80\\
S139 & \phn$-48.2$ & \nodata & \phn5.5 & \nodata & 11.3 & \nodata & O & \phn5.5 & 1.4 & \phn\phn$-2$\\
G106.250+00.975 & \phn$-53.8$ & \nodata & \phn6.0 & \nodata & 11.7 & \nodata & O & \phn6.0 & 1.3 & 102\\
S149 & \phn$-53.7$ & \nodata & \phn5.9 & \nodata & 11.8 & \nodata & O & \phn5.9 & 1.3 & $-108$\\
G112.029+01.050 & \phn$-41.3$ & \nodata & \phn4.4 & \nodata & 10.9 & \nodata & O & \phn4.4 & 1.3 & \phn79\\
S161 & \phn$-44.5$ & \nodata & \phn4.7 & \nodata & 11.1 & \nodata & O & \phn4.7 & 1.3 & \phn99\\
S173 & \phn$-31.2$ & \nodata & \phn3.1 & \nodata & 10.4 & \nodata & O & \phn3.1 & 1.2 & \phn$-47$\\
G122.774+02.855 & \phn$-56.0$ & \nodata & \phn5.7 & \nodata & 12.6 & \nodata & O & \phn5.7 & 1.2 & 285\\
S183 & \phn$-61.8$ & \nodata & \phn6.5 & \nodata & 13.2 & \nodata & O & \phn6.5 & 1.3 & 333\\
G125.072+01.641 & \phn$-83.5$ & \nodata & 10.1 & \nodata & 16.5 & \nodata & O & 10.1 & 1.8 & 288\\
S196 & \phn$-46.7$ & \nodata & \phn5.1 & \nodata & 12.7 & \nodata & O & \phn5.1 & 1.2 & 224\\
G136.742+00.471 & \phn$-52.0$ & \nodata & \phn5.9 & \nodata & 13.4 & \nodata & O & \phn5.9 & 1.4 & \phn48\\
G139.634+02.682 & \phn$-43.4$ & \nodata & \phn4.8 & \nodata & 12.6 & \nodata & O & \phn4.8 & 1.3 & 226\\
G149.740$-$00.202 & \phn$-71.2$ & \nodata & 17.4 & \nodata & 25.1 & \nodata & O & 17.4 & 5.4 & \phn$-61$\\
S217 & \phn$-21.9$ & \nodata & \phn3.8 & \nodata & 12.1 & \nodata & O & \phn3.8 & 1.8 & 219\\
G203.254$-$02.324 & \phn$29.9$ & \nodata & \phn5.1 & \nodata & 13.4 & \nodata & O & \phn5.1 & 2.0 & $-208$\\
G212.237$-$03.499 & \phn$58.2$ & \nodata & \phn9.8 & \nodata & 17.6 & \nodata & O & \phn9.8 & 2.6 & $-600$\\
G213.077$-$02.217 & \phn$55.5$ & \nodata & \phn8.7 & \nodata & 16.5 & \nodata & O & \phn8.7 & 2.3 & $-335$\\
G221.852$-$02.024 & \phn$38.0$ & \nodata & \phn4.0 & \nodata & 11.8 & \nodata & O & \phn4.0 & 1.2 & $-141$\\
S307 & \phn$38.1$ & \nodata & \phn3.7 & \nodata & 11.1 & \nodata & O & \phn3.7 & 1.2 & \phn53\\
G234.785$-$00.189 & \phn$38.6$ & \nodata & \phn3.7 & \nodata & 11.1 & \nodata & O & \phn3.7 & 1.3 & \phn$-12$\\

\enddata
\tablenotetext{a}{Kinematic distance ambiguity resolution: ``T''=tangent point distance; ``O''=outer Galaxy source}
%\tablecomments{Table~\ref{tab:distances} is available in its entirety in the electronic edition of the {\it Astrophysical Journal Supplement Series}. 
%A portion is shown here for guidance regarding its form and content.}
\label{tab:distances}
\end{deluxetable*}
%%%%%%%%%%%%%%%%%%%%%%%%%%%%%%%%%%%%%%%%%%%%%%%%%%
  
Kinematic distances use a model for Galactic rotation to derive
distances as a function of observed velocity for a given line of
sight.  Here we use the \citet{brand93} rotation curve.  Kinematic
distances are prone to large uncertainties in certain parts of the
Galaxy.  As in previous work, we estimate kinematic distance
uncertainties by adding in quadrature the uncertainties associated
with the rotation curve choice, streaming motions of 7\,\kms, and
changes to the Solar circular rotation speed.  Such uncertainties were
first computed by \citet{anderson12c}, and expanded to the entire
Galaxy by \citet{anderson14}.  We use the latter analysis here.

We do not compute kinematic distances for sources within $10\degree$
of the Galactic center or within $20\degree$ of the Galactic
anti-center because such distances would be uncertain by $\gtrsim
50\%$ \citep{anderson14}.  We also exclude \hii\ regions with multiple
RRL velocities for which the source velocity is unknown, as they have
at least two possible kinematic distances.  We exclude sources in the
first and fourth Galactic quadrants for which the absolute value of
the tangent point velocity is less than 10\,\kms.  Finally, we also do
not provide distances to regions whose distance uncertainties as
computed using the model of \citet{anderson14} are $>50\%$ that of
their kinematic distances.  
%These criteria are not
%mutually exclusive (e.g., a source within $10\degree$ of the Galactic
%center may also have multiple RRL velocities).

Sources in the inner Galaxy suffer from the well-known kinematic
distance ambiguity (KDA): inner Galaxy \hii\ regions have two possible
Heliocentric distances (called ``near'' and ``far'') for each measured
velocity.  The KDA does not exist for the 35 Outer Galaxy
\hii\ regions in our sample (including the seven first-quadrant
sources with negative RRL velocities and the seven fourth quadrant
sources with positive velocities).  For all these regions, there
is no ambiguity in the calculation of Galactocentric distances.

There are 75 inner-Galaxy single-velocity \hii\ regions in our sample.
The large regions detected here are too faint for
\hi\ Emission/Absorption analyses using extant \hi\ data
\citep[e.g.][]{anderson09a, anderson12c}.  Additionally, large
\hii\ regions have poor associations with molecular gas
\citep{anderson09b}, which makes using the \hi\ self-absorption
technique difficult to use.  As a result of these difficulties, we do
not attempt to resolve the kinematic distance ambiguity (KDA) for any
of the detected regions.  All of the kinematic distance ambiguity
resolutions (KDARs) are therefore for regions that have RRL velocities
within 10\,\kms\ of the tangent point velocity; we place these 11
  nebulae at the tangent point distance.  To determine distances to
the other 75 nebulae we would need more sensitive \hi\ observations.

\subsection{Extremely Luminous H\,{\bf\footnotesize II} Regions}
In Section~\ref{sec:intro} we mentioned the \hii\ region G52L, which
despite being extremely luminous was missed by many previous
\hii\ region surveys.  There are no equally luminous nebulae in the
present survey, although we lack distances and hence luminosities for
many sources.  We did, however, identify G039.515+00.511
(Figure~\ref{fig:example}), which has a VGPS 21\,cm continuum flux
density of $2.5\pm0.3$\,\jy and a {\it WISE}-identified angular
diameter of $17\arcmin$.  At a distance of 16.3\,\kpc, its physical
diameter is 80\,\pc.  Using the relationship in \citet{rubin68}, we
compute an ionizing photon emission rate of $10^{49.7}$\,s$^{-1}$,
which is 60\% that of G52L, and double that of the next most luminous
nebula in the present survey.  This is equivalent to the ionization
rate of one O4 star, or five O7 stars \citep{sternberg03}.  Although
not as luminous as G52L, this ionization rate places it in the top 1\%
of all Galactic \hii\ regions discovered to date (J.~Mascoop, 2017, in
prep.).

We detected the RRL emission from three additional nebulae with
{\it WISE}-defined angular diameters $>100\,\pc$: G070.099+01.629 (S99;
105\,\pc), G090.856+01.691 (115\,\pc), and G094.890-01.643 (130\,\pc).
The luminosities of these regions are much lower  that of
G039.515+00.511.  These are among the largest \hii\ regions known in
the Milky Way.  For example, there are only five other regions with
diameters $>100\,\pc$ in the {\it WISE} Catalog.

\section{Summary\label{sec:summary}}
We detected hydrogen RRL emission from 148 new large Galactic
\hii\ regions located north of $-42\degree$ declination ($266\degree >
\ell > -20\degree$ at $b = 0\degree$).  These regions were drawn from
the {\it WISE} Catalog of Galactic \hii\ Regions, and all have
infrared angular diameters $>260\arcsec$.  We also detect the helium
RRLs from 21 of the nebulae, and carbon RRLs from 16.

The regions are on average fainter than those of previous \hii\ region
surveys, larger in angular size, and have narrower RRL line widths.
We discover seven regions with line widths $<10\,\kms$, which implies
nebular electron temperatures $<1100\,\K$ if half the line width is
due to turbulence.  We do not have a satisfactory explanation for
these low line width sources. Previous authors have speculated
  that the narrow lines may be caused by interactions with nearby
molecular clouds or by partially ionized zones in the \hii\ region
photo-dissociation regions.  We note that all but one of the
narrow-line regions has three hydrogen RRL components, and all are
located toward the inner Galaxy.

We find one extremely luminous outer-Galaxy \hii\ region,
G039.515+00.511, which is one of the most luminous \hii\ regions known
in the Galaxy.  We also detect the RRL emission from three of the
largest known Galactic \hii\ regions, G070.099+01.629 (S99; 105\,\pc),
G090.856+01.691 (115\,\pc), and G094.890$-$01.643 (130\,\pc).

This survey completes the HRDS Northern census of Galactic
\hii\ regions.  The Northern HRDS surveys have together discovered 887
new Galactic \hii\ regions, whereas prior to the HRDS, the previously
known sample over the same area compiled in the {\it WISE} Catalog
numbered just 716.

\appendix
\section{Web Sites}
We have updated the GBT HRDS website with the results reported
here\footnote{http://go.nrao.edu/hrds}.  This site contains images
such as those in Figure~\ref{fig:example} for all detected sources, as
well as the same for all previous HRDS surveys.  We have also updated
the WISE Catalog of Galactic \hii\ Regions web
site\footnote{http://astro.phys.wvu.edu/wise} with results from these
observations.

%%%%%%%%%%%%%%%%%%%%%%%%%%%%%%%%%%%%%%%%%%%%%%%%%%
\begin{acknowledgments}
\nraoblurb\ This work is supported by NSF grant AST1516021 to LDA.  We
thank the staff at the Green Bank Telescope for their hospitality and
friendship during the observations and data reduction.  We thank West
Virginia University for its financial support of GBT operations, which
enabled some of the observations for this project.  Support for TVW
was provided by the NSF through the Grote Reber Fellowship Program
administered by Associated Universities, Inc./National Radio Astronomy
Observatory. Some of the targets for the present work were discovered
by Mr.~Root's students at Morgantown High School;
we thank them for their efforts.

\facility{Green Bank Telescope}
\software{TMBIDL \citep[][V8, Zenodo, doi:10.5281/zenodo.32790]{bania16}}
\end{acknowledgments}

\bibliographystyle{apj}
\bibliography{ref.bib}

\end{document}